\documentclass[%
 twocolumn,
superscriptaddress,
 amsmath,amssymb,
 aps,
 prx,
]{revtex4-2}

\usepackage{graphicx}
\usepackage{dcolumn}
\usepackage{bm}
\usepackage{microtype}
\usepackage{changes}
\usepackage{ulem}
\usepackage{enumerate}
\usepackage{amsmath,amssymb,amsfonts}
\usepackage{todonotes} 
\usepackage{natbib}
\usepackage{babel}
\usepackage{color}
\usepackage{soul}
\usepackage{comment}
\usepackage{mathtools,mathbbol,braket}
\usepackage{tikz}

\usetikzlibrary{arrows,shapes,positioning}

\newcommand{\fref}[1]{Fig.\hspace{0.025in}\ref{#1}}
\newcommand{\eref}[1]{Eq.\hspace{0.025in}(\ref{#1})}
\newcommand{\sref}[1]{Sec. \ref{#1}}

\newcommand{\bs}{\boldsymbol}

\begin{document}

\title{Conditional wavefunction theory: a unified treatment of molecular structure and nonadiabatic dynamics}

\author{Guillermo Albareda} \email{guillealpi@gmail.com}
\address{Nano-Bio Spectroscopy Group and European Theoretical Spectroscopy Facility (ETSF), Universidad del Pa\'is Vasco (UPV/EHU), Av. Tolosa 72, 20018 San Sebastian, Spain}
\address{Institute of Theoretical and Computational Chemistry, University of Barcelona, Mart\'i i Franqu\`es 1-11, 08028 Barcelona, Spain} 

\author{Kevin Lively} \email{kevin.lively@mpsd.mpg.de}
\address{Max Planck Institute for the Structure and Dynamics of Matter and Center for Free-Electron Laser  Science, Luruper Chaussee 149, 22761 Hamburg, Germany}
\address{The Hamburg Centre for Ultrafast Imaging, University of Hamburg, Luruper Chaussee 149, 22761, Hamburg, Germany}

\author{Shunsuke A. Sato} \email{ssato@ccs.tsukuba.ac.jp }
\address{Max Planck Institute for the Structure and Dynamics of Matter and Center for Free-Electron Laser  Science, Luruper Chaussee 149, 22761 Hamburg, Germany}
\address{Center for Computational Sciences, University of Tsukuba, 1-1-1 Tennodai, Tsukuba, Ibaraki, Japan}

\author{Aaron Kelly} \email{aaron.kelly@mpsd.mpg.de}
\address{Max Planck Institute for the Structure and Dynamics of Matter and Center for Free-Electron Laser  Science, Luruper Chaussee 149, 22761 Hamburg, Germany}
\address{The Hamburg Centre for Ultrafast Imaging, University of Hamburg, Luruper Chaussee 149, 22761, Hamburg, Germany}
\address{Department of Chemistry, Dalhousie University, Halifax B3H 4R2, Canada}

\author{Angel Rubio} \email{angel.rubio@mpsd.mpg.de}
\address{Nano-Bio Spectroscopy Group and European Theoretical Spectroscopy Facility (ETSF), Universidad del Pa\'is Vasco (UPV/EHU), Av. Tolosa 72, 20018 San Sebastian, Spain}
\address{Max Planck Institute for the Structure and Dynamics of Matter and Center for Free-Electron Laser  Science, Luruper Chaussee 149, 22761 Hamburg, Germany} 
\address{The Hamburg Centre for Ultrafast Imaging, University of Hamburg, Luruper Chaussee 149, 22761, Hamburg, Germany}
\address{Center for Computational Quantum Physics (CCQ), Flatiron Institute, 162 Fifth avenue, New York NY 10010, USA}

\date{\today}


\begin{abstract}
We demonstrate that a conditional wavefunction theory enables a unified and efficient treatment of the equilibrium structure and nonadiabatic dynamics of correlated electron-ion systems. The conditional decomposition of the many-body wavefunction formally recasts the full interacting wavefunction of a closed system as a set of lower dimensional (conditional) coupled `slices'. We formulate a variational wavefunction ansatz based on a set of conditional wavefunction slices, and demonstrate its accuracy by determining the structural and time-dependent response properties of the hydrogen molecule. We then extend this approach to include time-dependent conditional wavefunctions, and address paradigmatic nonequilibrium processes including strong-field molecular ionization, laser driven proton transfer, and Berry phase effects induced by a conical intersection. This work paves the road for the application of conditional wavefunction theory in equilibrium and out of equilibrium ab-initio molecular simulations of finite and extended systems.
\end{abstract}

\maketitle

\section{Introduction}\label{intro}

Emerging experimental capabilities in the precise manipulation of light and matter are opening up new possibilities to understand and exploit correlations and quantum effects that can be decisive in the functional properties of molecules and materials. 
Light-driven states can be not only designed to monitor and/or control the structure of molecules~\cite{mukamel1990femtosecond,zewail2000femtochemistry,corkum2007attosecond,sciaini2011femtosecond,blaga2012imaging,lepine2014attosecond,nisoli2017attosecond} and solids~\cite{devereaux2007inelastic,fink2013resonant,basov2017towards,buzzi2018probing,ruggenthaler2018quantum}, but also to form light–matter hybrid states with new physical properties~\cite{carusotto2013quantum,ebbesen2016hybrid,ribeiro2018polariton,hertzog2019strong,oka2019floquet,ozawa2019topological,rudner2020band,hubener2021engineering,genet2021inducing}.
In view of these exciting developments, accurate first principles theoretical techniques are also needed in order to help interpret observations, to enable the predictions of simplified models to be scrutinized, and ultimately, to help gain predictive control. Our ability to treat the full correlated quantum structure and dynamics of general electron-ion systems unfortunately remains limited by the unfavourable scaling of the many-body problem. 

A standard approach to address this problem in molecular and solid-state systems has been to `divide-and-conquer' in the sense that the electronic structure and the electron-nuclear interactions are treated separately. Introduced almost a century ago by Born and Oppenheimer~\cite{born1927quantentheorie}, the adiabatic approximation, i.e., the assumption that electrons adjust instantaneously to the motion of nuclei, is the cornerstone of this so-called standard approach. The Born-Oppenheimer (BO) approximation has been crucial to the development of a vast majority of approaches in quantum chemistry and condensed-matter theory~\cite{marx2009ab,ashcroft1976solid}, and the concept of ground state Born-Oppenheimer potential-energy surface (BOPES) is the foundation for understanding the properties of systems at thermal equilibrium such as chemical reactivity~\cite{eyring1931uber,miller1980reaction,heidrich2013reaction} and nuclear quantum effects~\cite{cha1989hydrogen,borgis1993dynamical,tuckerman1997quantum,raugei2003nuclear}, as well as of systems driven out of equilibrium~\cite{car1985unified,payne1992iterative,barnett1993born,zhang1998theory}. 

Accurately describing systems driven away from equilibrium and including nonadiabatic electron-nuclear effects, places even more stringent demands on the development of practical first principles tools. In the standard approach one directly builds upon the BO approximation by expanding the full molecular wavefunction in the Born-Huang basis~\cite{bornhuang}.
Within this framework, nonadiabatic processes can be viewed as nuclear wavepacket dynamics with contributions on several BOPESs, connected through nonadiabatic coupling terms that induce electronic transitions~\cite{domcke}.
In this picture, trajectory based quantum dynamics methods offer a trade-off between physical accuracy and computational cost~\cite{MillerPerspective,TullyPerspective,kapralreview}. Of these approaches, perhaps the most popular are Ehrenfest mean field theory~\cite{maclachlan}, and Tully's surface hopping dynamics~\cite{tully90}. Both of these approaches consist of an ensemble of uncorrelated trajectories. Reintroducing correlation, for example by using a variety of wavefunction ans\"atze~\cite{MCTDH_review,FMS,g-MCTDH,vmcg,MCE,cloning}, semiclassical techniques~\cite{scivr,fbivr}, the quantum-classical Liouville equation~\cite{qcle,mqcle,jfbts}, path-integral methods \cite{na-rpmd1,na-rpmd2}, or methods based on the exact factorization~\cite{exact_fac,ct-mqc,dish-xf}, allows for further accuracy with increased computational effort. 

While advances in \textit{ab initio} electronic structure theory in quantum chemistry and condensed matter have made computing the ground state energies both routinely efficient and rather accurate in many cases, obtaining accurate excited state information remains a challenging problem in its own right. Even in cases where the excited state electronic structure is available, performing fully quantum nuclear dynamics calculations using the standard approach quickly becomes infeasible ~\cite{zhang1998theory,beck2000multiconfiguration} as the memory required to store the information contained in the BOPESs grows rapidly with the number of correlated degrees freedom. In this respect, gaining the ability to rigorously treat selected nuclear degrees of freedom quantum mechanically without incurring an overwhelming computational cost is the goal.  

An alternative approach for describing quantum effects in coupled electron-ion systems is using a real space representation of all degrees of freedom. 
This route might sound less intuitive as it avoids routine concepts such as BOPESs and nonadiabatic couplings that are fundamental in the present description and understanding of quantum molecular dynamics. However, this feature might be turned into an attractive playground from the computational point of view, as these quantities are usually demanding to obtain and fit from ab initio electronic structure calculations.
In this framework, one of the leading approximate methods to describe the coupled electron-nuclear dynamics for large systems is time-dependent density functional theory coupled to classical nuclear trajectories through the Ehrenfest method~\cite{alonso2008efficient}. Due to its favourable system-size scaling, the real-space picture Ehrenfest method has been successful for a great many applications, from capturing phenomena associated with vibronic coupling in complex molecular systems~\cite{Lively_2020} and photodissociation dynamics in small molecules~\cite{castro2004excited}, to radiation damage in metals~\cite{mceniry2010modelling}; its efficiency allows calculations on large systems for even hundreds of femtoseconds~\cite{rozzi2013quantum}. It has also been recently combined with the nuclear-electronic orbital method as a way to include quantum effects for selected nuclear degrees of freedom, to study proton transfer processes in molecular excited states~\cite{zhao2021excited}. 

It is well-known, however, that the Ehrenfest approach can be inaccurate due to its mean-field nature. One classic example of this breakdown occurs in photochemical reaction dynamics, where mean field theory can often fail to correctly describe the product branching ratios~\cite{Tully1990,TullyPerspective}. Generally speaking, the mean field description of any transport property can potentially suffer some deficiency; this is sometimes referred to as a violation of detailed balance~\cite{parandekar2005mixed}, but it ultimately stems from the lack of time-translational invariance that is inherent to any approximate method that does not rigorously preserve the quantum Boltzmann distribution~\cite{jacobi_qcle}. 

The conditional wavefunction (CWF) framework introduced in Ref.~\cite{Albareda2014} offers a route to go beyond the limits of mean field theory while retaining a real-space picture; it is an exact decomposition and recasting of the total wavefunction of a closed quantum system~\cite{Bohm_measure}. When applied to the time-dependent Schr\"odinger equation the conditional decomposition yields a set of coupled, non-Hermitian, equations of motion~\cite{Albareda2014}. One can draw connections between CWF theory and other formally exact frameworks proposed in the literature to develop novel approximate schemes that provide a completely new perspective to deal with the long-standing problems of nonadiabatic dynamics of complex interacting systems~\cite{Albareda2015,Albareda2016}. An example is the time-dependent interacting conditional wavefunction approach (ICWF)~\cite{Albareda2019,albareda2020bohmian}, a recently introduced method for performing quantum dynamics simulations that is multi-configurational by construction. Using a stochastic wave-function ansatz that is based on a set of interacting single-particle CWFs, the ICWF method is a parallelizable technique which achieves quantitative accuracy for situations in which mean-field theory drastically fails to capture qualitative aspects of the dynamics, such as quantum decoherence, using orders of magnitude fewer trajectories than the converged mean-field results~\cite{Albareda2019}. 

In this work we introduce an exact time-independent version of the CWF mathematical framework. The time-independent CWF framework is formulated in real-space, and it is an exact decomposition of the time-independent wavefunction of a closed quantum system that yields a set of of coupled nonlinear eigenvalue problems and associated conditional eigenstates.  Based on this framework, we put forth a static-basis version of the ICWF method, which allows us to establish an efficient and accurate algorithm for calculating the ground- and excited-state structure of correlated electron-nuclear systems and eventually extended systems. Importantly, the combination of the static version of the ICWF method using a time-dependent conditional eigenstate basis sets the stage for the implementation of a general purpose \textit{ab initio} molecular simulator that is formulated in the real-space picture and that self-consistently treats stationary states as well as driven dynamics.

This manuscript has the following structure: in ~\sref{TICWF} we define the mathematical structure of the time-independent version of the CWF framework. Based on these results we put forth an imaginary time version of the ICWF technique in \sref{var_TISE} for solving the time-independent Schr\"odinger equation and the performance of the resulting algorithm is assessed through the calculation of the ground-state and the low-lying excited state BOPESs of the hydrogen molecule in one-dimension. In \sref{var_TDSE} a real-time extension of this multiconfigurational ansatz is presented, along with an algorithm for solving the time-dependent Schr\"odinger equation using a stochastic static-basis ansatz. The ability of the resulting algorithm in capturing static and dynamic properties is then assessed by evaluating the absorption spectrum and a laser-induced dynamics of the aforementioned H$_2$ model system. In \sref{TDCWF} we revisit the exact time-dependent CWF framework and in \sref{nvar_TDSE} we present the dynamical ICWF approach to the time-dependent Schr\"odinger equation. The performance of the time-dependent ICWF method in combination with its imaginary time variation for preparing the initial state is demonstrated for three model systems, viz., a laser driven proton-coupled electron transfer model, an electron-atom scattering process, and an example of the effect of the Berry phase in the quantum nuclear dynamics through a conical intersection. A summary of the main results of this work and an outlook on future directions is offered in \sref{CONCL}.

\section{Conditional Eigenstates}\label{TICWF}

We begin by considering a closed system with $n$ electrons and $N$ nuclei, collectively denoted by $\bs x = (\bs r,\bs R)$. We use the position representation for both subsets; lowercase symbols will be used for the electronic subsystem, e.g., $\bs r = \{ \bs r_1 s_1,...,\bs r_{n}  s_n\}$, and uppercase symbols $\mathbf{R} = \{ \mathbf{R}_1 \sigma_1,...,\mathbf{R}_{N} \sigma_N\}$ for the nuclear subsystem. Hereafter, electronic and nuclear spin indices, respectively $ s_j$ and $ \sigma_j$, will be made implicit for notational simplicity, and, unless otherwise stated, all expressions will be expressed in atomic units.

\begin{figure*}
\includegraphics[width=\textwidth]{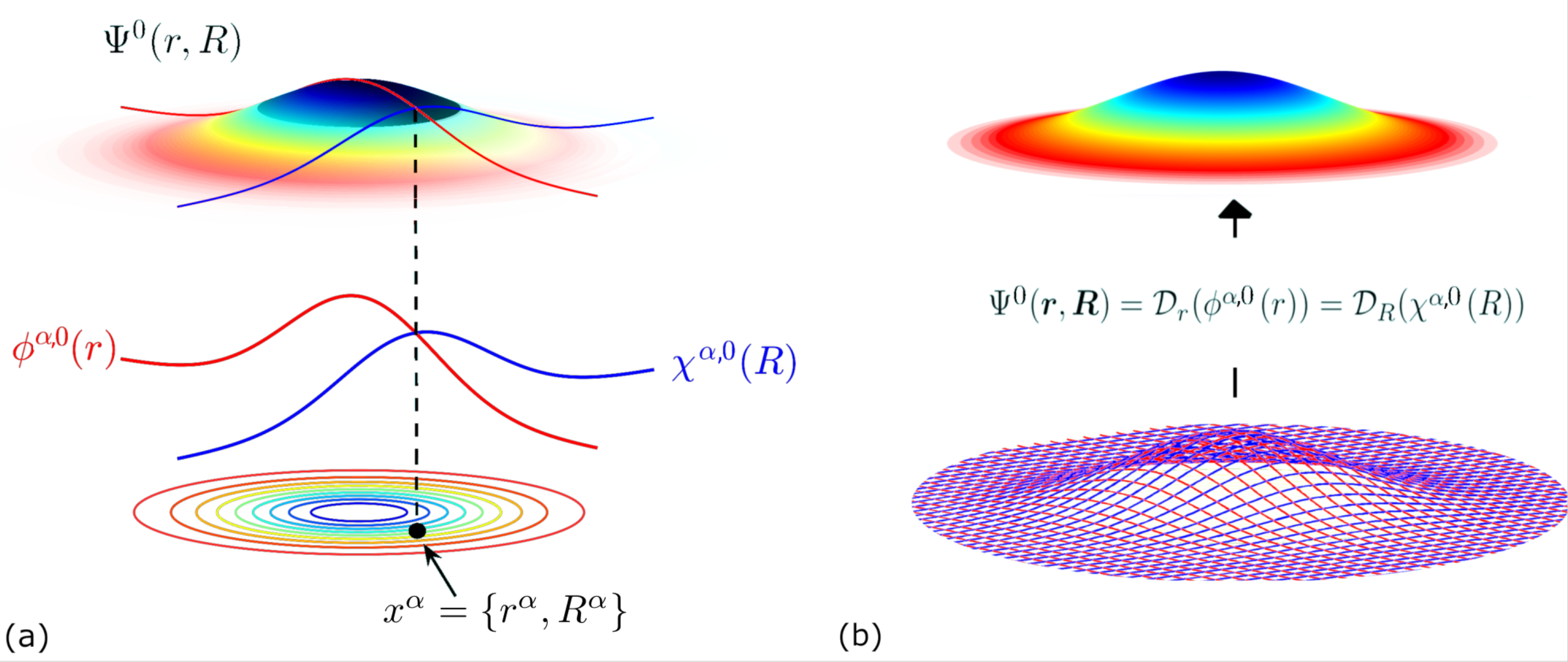}
\caption{Schematic representation of the CWF approach to the time-independent Schr\"odinger equation for one electron and one nucleus in one dimension, i.e., $\bs x = (r, R)$. (a) The full ground state $\Psi^0( r,  R)$ is plotted together with a pair of conditional ground states $\phi^{\alpha,0}(r)$ (in red) and $\chi^{\alpha,0}(R)$ (in blue) for a given position of the full configuration space $\{r^{\alpha}, R^{\alpha}\}$. Contour plots of the molecular wavefunction are also shown for clarity. (b) The exact solution of the time-independent Shcr\"odinger equation in \eref{eq1:1} can be reconstructed provided a sufficiently large ensemble of sampling points $x^{\alpha} = \{{r}^{\alpha},{R}^{\alpha}\}$. This can done by applying the reassembling transformation $\mathcal{D}_{{r}}$ or $\mathcal{D}_{{R}}$ to the ensemble of conditional eigenstates $\phi^{\alpha,0}({r})$ or $\chi^{\alpha,0}_A({R})$ respectively.}
\label{CWFscheme_ti}
\end{figure*}

The time-independent CWF can be constructed starting from the non-relativistic time-independent Schr\"odinger equation in position representation,
\begin{equation}\label{eq1:1}
    \hat{H} \Psi^\gamma(\bs x) = E^\gamma \Psi^\gamma(\bs x),
\end{equation}
where $\Psi^\gamma(\bs x)$ is an eigenstate of the molecular Hamiltonian $\hat{H}$ with label $\gamma$, and corresponding energy eigenvalue $E^{\gamma}$. The molecular Hamiltonian operator $\hat{H}$ in \eref{eq1:1} can be written as:
\begin{equation}\label{Hamiltonian}
    \hat H = \sum_{j=1}^{N\times n} \hat{T}_j(\bs x_j) + W(\bs x),
\end{equation}
where the kinetic energy operators are $\hat{T}_j=\frac{1}{2m_j}(-i\hbar \nabla_j - z_j \textbf{A}(\mathbf{x}_j))^2$, being $m_j$ and $z_j$ the characteristic mass and charge of particle $j$ respectively. The full electron-nuclear potential energy of the system is $W(\bs x)$ (written in the position basis rather than, say, the BO or Born-Huang basis), and $\textbf{A}$ is the vector potential due to an arbitrary static external electromagnetic field.

Note that the total Hamiltonian in \eref{eq1:1} is invariant under translations and rotations of all particles. This means that the eigenstates of the system will be invariant under transformations by the translation and rotation group. Together with the inversion symmetry, this implies that all one-body quantities such as the electron density or any nuclear reduced density are constant and that two-particle position correlation functions only depend on the distance between their arguments. This is obviously not a convenient starting point to describe the structure of a quantum system. The solution to this problem relies on transforming the Hamiltonian to a fixed coordinate system that reflects the internal properties of the system~\footnote{This has been discussed at length in the literature. A general and very elegant discussion on the various ways the body-fixed frame can be chosen is given in Refs.~\cite{littlejohn1997gauge,kreibich2008multicomponent}}. This is, in general, not a trivial task and hereafter we will assume that \eref{eq1:1} already reflects such  internal properties, either by exploiting a particular symmetry of the system or by simply introducing a parametric dependence on, e.g., a fixed (heavy) nuclear position.

At this point we can decompose the eigenstates $\Psi^\gamma(\bs x)$ in terms of single-particle conditional eigenstates of either of the two subsystems, which are defined as follows:
\begin{align}\label{CES1}
	& \psi_{i}^{\alpha,\gamma}(\bs x_i) :=
 \int d\bar{\bs x}_i \delta(\bar{\bs x}_i^{\alpha} - \bar{\bs x}_i) \Psi^\gamma.
\end{align} 
Here the index $\alpha$ denotes the particular conditional slice and $\bar{\bs x}_i = (\bs x_1,...,\bs x_{i-1},\bs x_{i+1},...,\bs x_{n\times N})$ are the coordinates of all degrees of freedom in the system except $\bs{x}_i$. Similarly, $\bar{\bs x}_i^{\alpha} = (\bs x_1^{\alpha},...,\bs x_{i-1}^{\alpha},\bs x_{i+1}^{\alpha},...,\bs x_{n\times N}^{\alpha})$ are some particular positions of all system degrees of freedom except $\bs x_i$.  As shown schematically in \fref{CWFscheme_ti}, the conditional eigenstates in \eref{CES1} represent one-body slices of the full many-body eigenstates $\Psi^\gamma(\bs x)$ taken along the coordinate of the $i$-th degree of freedom.
The particle placement $\bs x^\alpha$ defining the CWFs has not yet been specified, and although in principle it can be chosen arbitrarily, it will be proven convenient in practice to exploit importance sampling techniques.

Evaluating \eref{eq1:1} at $\bar{\bs x}_i^{\alpha}$ by applying the integral operator in \eref{CES1} yields conditional eigenstates that are the solutions of the following eigenvalue problem:
\begin{align}\label{eigs_e}
   & \left(\hat{T}_i + W^{\alpha}_{i} + \eta_{i}^{\alpha,\gamma}\right) \psi_{i}^{\alpha,\gamma} = E^\gamma \psi_{i}^{\alpha,\gamma},
\end{align}
where we introduced $W^{\alpha}_i(\bs x_i) = W(\bs x_i,\bar{\bs x}_i^{\alpha})$, with $W(\bs x)$ the full electron-nuclear interaction potential appearing in the Hamiltonian of \eref{Hamiltonian}. 
In addition, $\eta_{i}^{\alpha,\gamma}(\bs x_i)$ are kinetic correlation potentials given by, 
\begin{align}\label{CP1}
  &\eta_{i}^{\alpha,\gamma}(\bs x_i)  = \sum_{j\neq i}^{n\times N} \frac{\hat T_j \Psi^\gamma}{\Psi^\gamma} \bigg|_{\bar{\bs x}_i^{\alpha}}.
\end{align}

Provided a large enough collection of CWFs satisfying \eref{eigs_e}, an exact solution of \eref{eq1:1} can be reconstructed by undoing the conditional decomposition of \eref{CES1} (see \fref{CWFscheme_ti}.b)~\cite{Albareda2014}. That is, given a set of conditional slices that sufficiently spans the support of $\Psi^\gamma$, then the corresponding conditional eigenstates can be used to reassemble the full electron-nuclear wavefunction, 
\begin{equation}\label{reconstruction_ti}
    \Psi^\gamma(\bs x) = \mathcal{D}_{\bs x_i}\left(\psi_{i}^{\alpha,\gamma}\right),
\end{equation} 
using the transformations $\mathcal{D}_{\bs x_i}$ which are discussed in more detail in Appendix~\ref{sec:reconst}. This expression, \eref{reconstruction_ti}, can be used to evaluate the kinetic correlation potentials in \eref{CP1}. In this way, the generalized one-body eigenvalue problem in \eref{eigs_e} can be understood as an exact decomposition and recasting of the eigensolution of the full electron-nuclear system, that yields a set of coupled, non-Hermitian, eigenvalue problems.

\subsection{Time-independent Hermitian approximation}\label{ti_hermitian_sec}

An approximate solution to \eref{eigs_e} can be formulated by expanding the kinetic correlation potentials around the sampling coordinates $\bs x^{\alpha}$ using Taylor series, and then truncating at zeroth order, i.e.:
\begin{align}\label{hermitian_e}
    &\eta_{i}^{\alpha,\gamma} (\bs x_i) \approx f(\bar{\bs x}_i^{\alpha}).
\end{align}
At this level the kinetic correlation potentials engender only a global phase that can be simply omitted as expectation values are invariant under such global phase transformations. Note that these approximated kinetic correlation potentials can be alternatively obtained by introducing a mean field ansatz $\Psi^\gamma (\bs x) = \prod_{i=1}^{n\times N} \psi(\bs x_i)$ into \eref{CP1}. By making this approximation the eigenvalue problems in \eref{eigs_e} are restored to a Hermitian form,
\begin{align}\label{h_eigs_e}
   & \left(\hat{T}_i + W^{\alpha}_{i} \right)\psi_{i}^{\alpha,\gamma} \approx E^\gamma \psi_{i}^{\alpha,\gamma}.
\end{align}
The Hermitian limit allows the full many-body problem to be approximated as a set of independent single-particle problems. That is, the superscript $\gamma$ refers exclusively to the conditional eigenstate excitation number.

\section{Static properties with conditional eigenstates}\label{var_TISE}

In general the higher order terms in the Taylor expansion of the kinetic correlation potentials are non-negligible. However, one can still take advantage of the simple Hermitian form of the conditional eigenvectors (hereafter referred to as conditional wavefunctions CWFs) in \eref{h_eigs_e} to design an efficient many-body eigensolver by utilizing them as bases for electronic and nuclear degrees of freedom in a variational wavefunction ansatz.

While there is a diverse literature spanning decades on different forms for variational electron-nuclear wavefunction ans\"{a}tze, for illustrative (and practical) purposes we employ a sum-of-products form, which in the language of tensor decompositions is referred to as the canonical format~\cite{C5CP01215E}. For each degree of freedom $\bs x_i$ we utilize a given electronic or nuclear CWF, respectively, coming from solutions to ~\eref{h_eigs_e}, in order to approximate the $\gamma^{\text{th}}$ full system exact excited state as follows:
\begin{align}
    \Psi^\gamma(\bs x) = \sum_{(\lambda,\nu)=(1,1)}^{(\text{N}_c,\text{M})} C_{\lambda,\nu}^\gamma\prod_{i=1}^{n\times N}\psi_{i}^{\lambda,\nu}(\bs x_i),\nonumber\\
    =\sum_{\alpha=1}^{\text{N}_c\text{M}}  C_{\alpha}^\gamma\prod_{i=1}^{n\times N}\psi_{i}^{\alpha}(\bs x_i).
\label{eq3:1}
\end{align}
Where in the second line we have rearranged the sum over particle position $\lambda\in\{1,\ldots, \text{N}_c\}$ and excited CWF $\nu\in\{1,\ldots,\text{M}\}$ into a single index $\alpha=\lambda+N_c(\nu-1)$, such that $\alpha\in\{1,\ldots,\text{N}_c\text{M}\}$.
The particle placement $\bs x^{\alpha}$ defining the conditional potentials $W_{i}^{\alpha}$ has not yet been specified, and in principle it can be chosen arbitrarily, however in practice we choose to sample from initial guesses for the reduced densities of the electronic and nuclear subsystems.

We refer to this ansatz (Eq. \ref{eq3:1}) as being in canonical format because we do not mix all possible CWFs $\psi_i^{\lambda,\nu}$ for all possible degrees of freedom $\bs x_i$, as one does with a single particle function bases across the different system degrees of freedom in the Tucker format employed in the Multi-Configurational Time-Dependent Hartree (Fock) -- MCTDH(F) ~\cite{beck2000multiconfiguration,MCTDH_review}, and Multi-Configurational Electron Nuclear Dynamics ans\"{a}tze~\cite{ulusoy2012multi}. In principle this choice can be relaxed, and one can utilize various choices of tensor network representation for the expansion coefficients $\bf{C}$, such as Matrix Product States or hierarchichal Tucker formats, which when employed in the Multi-Layer extension~\cite{Vendrell2011,manthe2008multilayer} of MCTDH allow for an increase in efficiency for certain problems. However, since the time dependence of the ansatz in \eref{eq3:1} is entirely within the expansion coefficients, one only needs to calculate the matrix elements at time zero, creating a quite efficient time propagation framework. Note that although we use a simple Hartree product over electronic degrees of freedom, the above ansatz can be straightforwardly extended to have fermionic anti-symmetry via treating the CWFs as the spatial component of spin orbitals in Slater determinants.

Hereafter, and for reasons that will be apparent later, we will call \eref{eq3:1} the static-basis ICWF (or sta-ICWF) ansatz.
With this ansatz in hand, we then consider a solution of \eref{eq1:1} based on the imaginary time propagation technique~\cite{kosloff1986direct}, i.e.:
\begin{equation}
    \frac{d}{d\tau} \Psi^\gamma(\bs x,\tau) = - \hat{H}^\gamma \Psi^\gamma(\bs x,\tau),
\label{eq5:1}    
\end{equation}
where 
\begin{equation}\label{eq5:2}    
    \hat{H}^\gamma(\bs x) = \left( \mathbb{I} - \sum_{\zeta=1}^{\gamma-1} \hat{P}^\zeta \right) \hat{H}(\bs x) \left( \mathbb{I} - \sum_{\zeta=1}^{\gamma-1} \hat{P}^\zeta \right) ,
\end{equation}
and $\hat{P}^\zeta = \Psi^\zeta \Psi^{\zeta\dagger}$ are projectors used to remove the wavefunctions $\Psi^\zeta$ from the Hilbert space spanned by $\hat{H}$.
The first excited state, for instance, is thus obtained by removing the ground state from the Hilbert space, which makes the
first excited state the ground state of the new Hamiltonian, $\hat{H}^{\zeta=1}$.

By introducing the ICWF ansatz of \eref{eq3:1} into \eref{eq5:1} we find an equation of motion for the coefficients $\mathbf{C}^\gamma = \{C_{1}^\gamma,\ldots ,C_{\text{N}_c\text{M}}^\gamma\}$:
\begin{align}\label{eq5:3}    
    \frac{{d\mathbf{C}^\gamma}}{{d\tau}} = - \mathbb{S}^{-1} \mathbb{H} \mathbf{C}^\gamma(\tau) + \mathbb{S}^{-1} \sum_{\xi=1}^{\gamma-1} \left( \mathbb{H} \mathbb{C}^\xi \mathbb{S} + \mathbb{S}\mathbb{C}^\xi\mathbb{H} \right) \mathbf{C}^\gamma(\tau)   \nonumber \\
    - \mathbb{S}^{-1} \sum_{\xi=1}^{\gamma-1} \sum_{\nu=1}^{\gamma-1} \mathbb{S}\mathbb{C}^\xi \mathbb{H} \mathbb{C}^\nu \mathbb{S} \mathbf{C}^\gamma(\tau)
\end{align}
where $\mathbb{C}^\xi = \mathbf{C}^{\xi} \mathbf{C}^{\xi,\dagger}$, and the matrix elements of $\mathbb{H}$ and $\mathbb{S}$ are:
\begin{align}\label{eq5:4}    
    &\mathbb{S}_{\alpha\beta} = 
    \prod_{i=1}^{n\times N}\int d\bs x_i \psi_{i}^{\alpha *}\psi_{i}^{\beta},
\\  \label{eq5:4_bis}
    &\mathbb{H}_{\alpha\beta} = \prod_{i=1}^{n\times N}
    \int d\bs x_i \psi_{i}^{\alpha *}\hat{H}\psi_{i}^{\beta},
\end{align}
where again, the $\alpha,\beta$ indices refer to the index over particle placement and excited CWFs. 
Obtaining these matrix elements involves a sum over all two-body interactions across each degree of freedom, and a sum across one-body operators. In practice $\mathbb{S}$ may be nearly singular, but its inverse can be approximated by the Moore-Penrose pseudoinverse. 

Based on solving the system of equations in \eref{eq5:3} for $\mathbf{C}^\gamma$, one already has the ingredients to put forth a time-independent ICWF eigensolver algorithm that will ultimately be used to evaluate the expectation value of generic observables $\hat{O}(\bs x)$. Given an approximate solution to the eigenfunction $\Psi^\gamma(\bs x)$, the expectation value of $\mathbb{O}$ reads:
\begin{equation}\label{equil_expectation}
    \langle \hat{O}\rangle_\gamma = \mathbf{C}^{\gamma\dagger} \mathbb{O}\mathbf{C}^\gamma,
\end{equation}
with the matrix elements of $\mathbb{O}$ being given by an analogous expression to \eref{eq5:4_bis}.

\subsection{Example I: Ground and Excited BOPESs of H$_2$}\label{example_BO}

As an illustrative example we now calculate the BOPESs of a model for the H$_2$ molecule. We adopt a model where the motion of all particles is restricted to one spatial dimension, and the center-of-mass motion of the molecule can be separated off~\cite{javanainen1988numerical}. In this model the relevant coordinates are the internuclear separation, $R$, and the electronic coordinates, $r_1$ and $r_2$. The Hamiltonian, written in terms of these coordinates, is
\begin{align}\label{H_H2}
    H(r_1,r_2,R) = -\frac{1}{\mu_n}\frac{\partial^2}{\partial R^2} + \frac{1}{R} + W_{ee}(r_1,r_2) 
    \nonumber\\
    + \sum_{i=1}^2\left( -\frac{1}{2\mu_e}\frac{\partial^2}{\partial r_i^2} + W_{en}(r_i,R) \right),
\end{align}
where $\mu_n=M/2$ and $\mu_e = M/(2M+1)$ are the reduced nuclear and electronic masses respectively, and $M$ is the proton mass. In \eref{H_H2} the electron-electron repulsion and the electron-nuclear interaction are represented by soft-Coulomb potentials,
\begin{align}
   & W_{ee}(r_1,r_2) = \frac{1}{\sqrt{(r_1-r_2)^2+\epsilon_{ee}}},
    \\
   & W_{en}(r,R) = -\frac{1}{\sqrt{(r-R/2)^2+\epsilon_{en}}}
                  -\frac{1}{\sqrt{(r+R/2)^2+\epsilon_{en}}},
\end{align}
i.e., the Coulomb singularities are removed by introducing a
smoothing parameter $\epsilon_{ee} = 2$ and $\epsilon_{en} = 1$. 
The above model system qualitatively reproduces all important
strong-field effects such as multiphoton ionization, above threshold ionization, or high-harmonic generation~\cite{eberly1989nonlinear,su1991model,schwengelbeck1994ionization}. Moreover, it has provided valuable information in the investigation of electron correlation effects~\cite{lein2000intense,bauer1997two,lappas1998electron}. 


For this model, the BOPESs are defined by the following electronic eigenvalue problem:
\begin{equation}
{\mathcal{H}}_{el}(r_1,r_2; R) \Phi^\gamma(r_1,r_2; R) = \epsilon^\gamma(R)\Phi^\gamma(r_1,r_2; R),
\label{eq:BO2}
\end{equation}
where $\mathcal{H}_{el} = \hat{H}-\hat{T}_{nuc}$, and $\left\{\Phi^\gamma(r_1,r_2; R)\right\}$ are the (complete, orthonormal) set of BO electronic states. A parametric dependence on the nuclear coordinates is denoted by the semicolon in the argument. The BOPESs, $\epsilon^\gamma(R)$, can be calculated using the imaginary time sta-ICWF method described in Eqs.~\eqref{eq5:1}-\eqref{eq5:4_bis} along with a simplified version of the ansatz in Eq.~\ref{eq3:1} that is specialized to this particular case of parametric nuclear dependence. A thorough description of the numerical procedure as well as the convergence behaviour of the sta-ICWF method for this model can be found in Appendix~\ref{SI_BOPES}.

In \fref{BOPES_all} we show the first five BOPESs calculated via the sta-ICWF approach using $(\text{N}_c,\text{M}) = (32,5)$. In the top panel, the exact BOPESs are plotted against the sta-ICWF data, overlaid as solid gold lines. The results demonstrate that the sta-ICWF ansatz used in a variational context captures the entire group of excited BOPES landscape over this energy range. As a point of comparison, in the bottom panel of \fref{BOPES_all} we also show the result of mean-field type calculations of the BOPESs for this system. Specifically, we show Hartree-Fock and configuration interaction singles (CIS) data for the ground state and excited state BOPESs respectively, which suffer from well-known inaccuracies in capturing the binding energy and excited state properties of the system.
\begin{figure}
\includegraphics[width=\columnwidth]{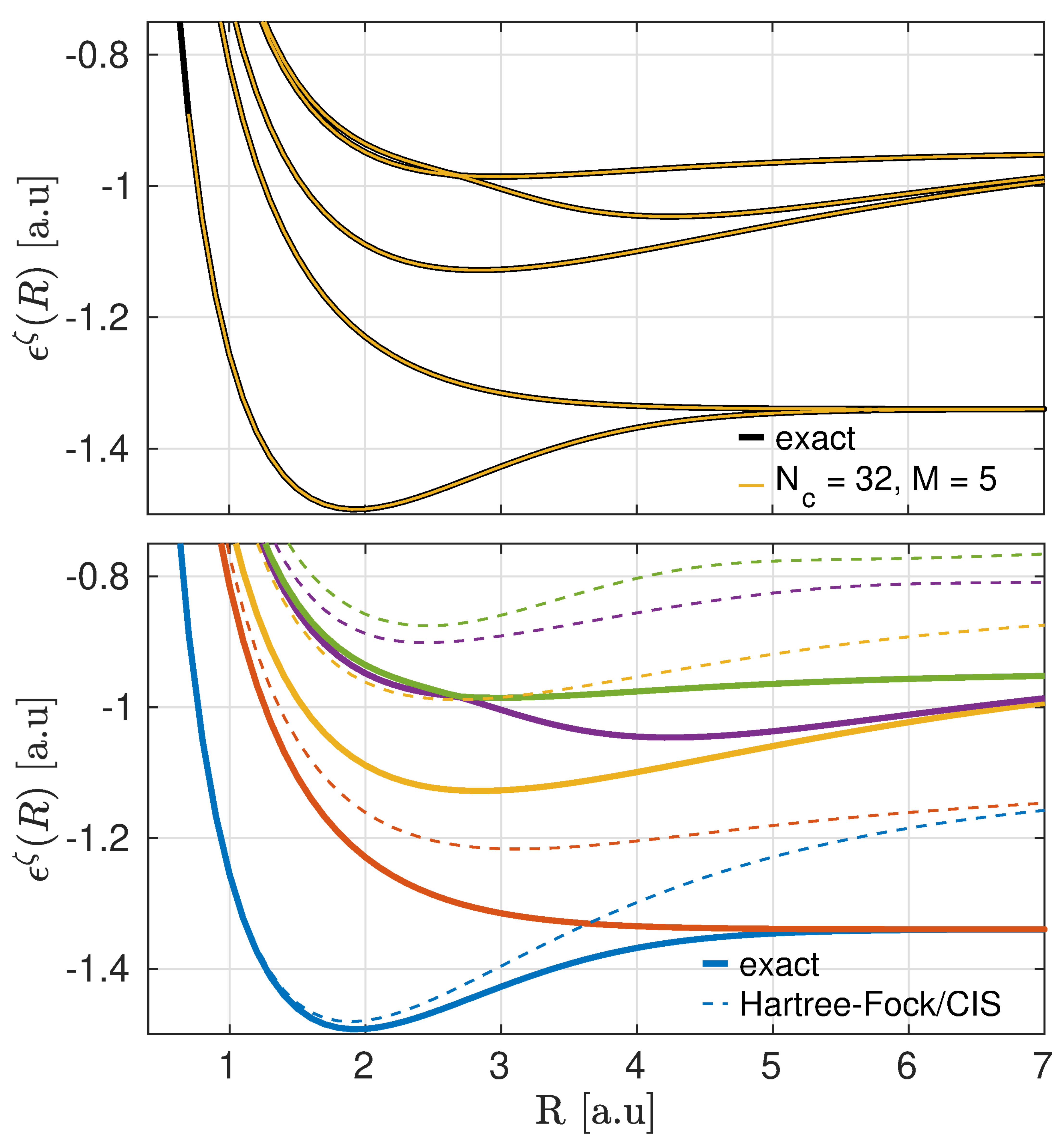}
\caption{Exact first five BOPESs of the one-dimensional H$_2$ model system (solid black lines). sta-ICWF results for $(\text{N}_c,\text{M}) = (32,5)$ are shown in the top panel (solid gold lines). Hartree-Fock and CIS results for the ground state and excited state BOPESs respectively are shown in the bottom panel (dashed lines) alongside exact results (solid lines) and color coordinated via calculated excited states.}
\label{BOPES_all}
\end{figure}

\section{Time-dependent properties with conditional eigenstates}\label{var_TDSE}
The sta-ICWF eigensolver described above can be easily extended to describe dynamical properties. For that, we consider the time-dependent Schr\"odinger equation:
\begin{equation}
    i\frac{d}{dt} \Psi(\bs x,t) = \hat{H}(t) \Psi(\bs x,t),
\label{TDSE}
\end{equation}
where $\Psi(\bs x, t)$ is the electron-nuclear time-dependent wavefunction and the Hamiltonian of the system $\hat H(t)$ may contain a time-dependent external electromagnetic field.

In practice, we are interested in situations where the initial wavefunction is the correlated electron-nuclear ground state, i.e., $\Psi(\bs x,0) = \Psi^{\gamma=0}(\bs x)$, and some nonequilibrium dynamics is triggered by the action of an external driving field. We can then decompose the time-dependent many-body wavefunction as in \eref{eq3:1} by restricting it to the case of $\gamma = 0$ (hereafter we omit the superscript $\gamma$ for clarity).  
The expansion coefficients $C_{\alpha}$ then obey an equation of motion that can be obtained either by inserting \eref{eq3:1} directly into \eref{TDSE} or by utilizing the Dirac-Frenkel variational procedure~\footnote{Both procedures lead to identical equations of motion due to the only time dependent variational parameter being the expansion coefficients}:
\begin{equation}\label{eq: ICWF prop}
    \frac{d}{dt} \bs C (t) = -i\mathbb{S}^{-1}\mathbb{H}(t)\bs{C}(t).
\end{equation}
In \eref{eq: ICWF prop}, the matrix elements of $\mathbb{S}$ and $\mathbb{H}$ are identical to the ones defined in Eqs.~\eqref{eq5:4} and \eqref{eq5:4_bis}, with the hamiltonian's time dependence coming from any external fields. The values of the coefficients at time $t=0$, i.e., $\bs C(0)$, may be obtained from the imaginary time sta-ICWF method of \eref{eq5:3}.
In this way, the combination of the imaginary time and real-time sta-ICWF methods yield a ``closed-loop'' algorithm for the structure and dynamics of molecular systems that does requires explicit BO state information as an input to the method. 


\subsection{Example II: Optical absorption spectrum of H$_2$}
\label{example_abs}

\begin{figure}
\includegraphics[width=\columnwidth]{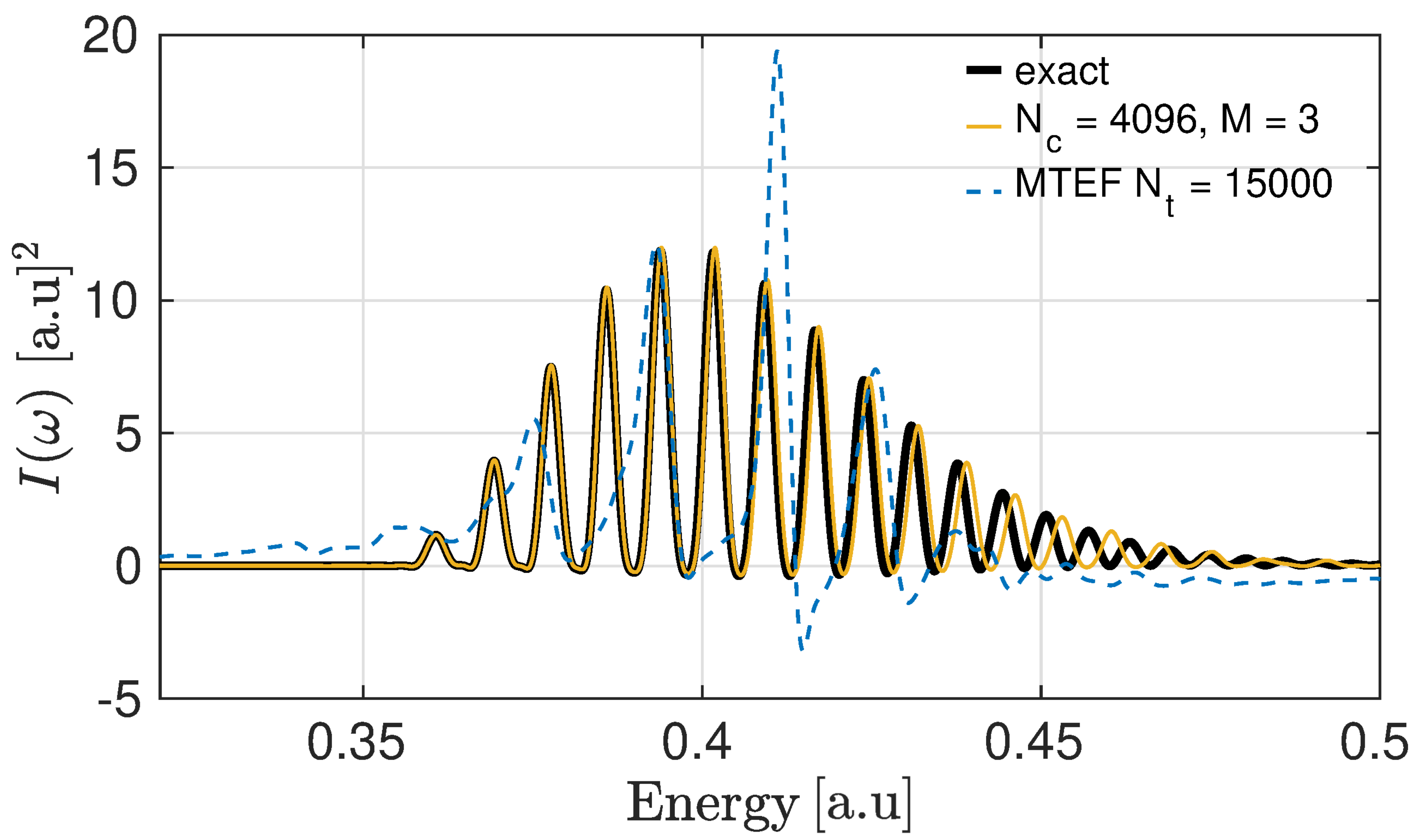}
\caption{S$_2$ $\leftarrow$ S$_0$ spectra of ICWF-Kick (gold) and MTEF-Kick (blue), compared to the exact peaks placement overlaid as black line, showing that while mean-field theory is unable to capture the qualitatively the correct vibronic lineshape spacing and intensity, the sta-ICWF approach accurately captures the exact spectrum.}
\label{fig:ICWF absorbtion}
\end{figure}

Here we demonstrate an application of the real-time sta-ICWF approach to simulate the optical absorption spectrum for the molecular Hydrogen model introduced in \sref{example_BO}. We utilize the ``$\delta$-kick" method of Yabana and Bertsch~\cite{Yabana1996}, where an instantaneous electric field $E(t)=\kappa\delta(t)$, with perturbative strength $\kappa\ll 1$a.u.$^{-1}$ couples to the dipole moment operator $\mu = r_1 + r_2$, and thereby produces an instantaneous excitation of the electronic system to all transition dipole allowed states. The resulting (linear) absorption spectra can then calculated via the dipole response, $\Delta\mu(t) = \mu(t) - \mu(0^-)$
\begin{equation}\label{absorption}
    I(\omega) = \frac{4\pi\omega}{c\kappa}\Im\left[\int_{0}^{\infty}  e^{i\omega t} \braket{\Delta\mu(t)} dt \right].
\end{equation} 
In practice, due to the finite time propagation, the integrand is also multiplied by a mask function $\mathcal{M}(t)$ that smoothly vanishes at the final simulation time $T_f$.

The system is first prepared in the ground state using the imaginary time sta-ICWF. See Appendix~\ref{SI_GS} for a thorough description of the imaginary-time sta-ICWF method and its use for preparing the ground state the H$_2$ model system. The field-driven dynamics is then generated by applying the kick operator to the relevant degree of freedom. A thorough description of the numerical procedure as well as the convergence behaviour of the sta-ICWF method for this model can be found in Appendix~\ref{SI_optical}.

For the H$_2$ model the occupation of excited electronic states and subsequent coupled electron-nuclear dynamics produce a characteristic vibronic peak structure usually explained via the Franck-Condon vertical transition theory.
In the top panel of Fig. \ref{fig:ICWF absorbtion} we show vibronic spectra calculated both with sta-ICWF for the absorption from $S_0$ to $S_2$ in comparison with the numerically exact results, also calculated via the $\delta$-kick approach.
For sta-ICWF, we found that $\text{N}_c=4096$ and $\text{M}=3$ was sufficient to obtain accurate results. The results demonstrate that the sta-ICWF ansatz used in a variational context achieves an accurate vibronic spacing, and furthremore it captures not only the electron-nuclear correlation inherent to vibronic spectra, but also solves the electron-electron subsystem accurately. The deviation from the exact results does grow with increasing energy, although this is ameliorated with increasing $\text{N}_c$ and $\text{M}$, and can in principle be eliminated at large enough values of these parameters (see Appendix~\ref{SI_optical}). 

For comparison, we show also mean-field results for the vibronic spectra. Specifically, we calculated the absorption spectrum with the multi-trajectory Ehrenfest $\delta$-kick method (MTEF-kick)~\cite{Lively_2020}, overlaid as dashed blue lines. We see that the vibronic spacing calculated with the MTEF-kick approach fails in capturing the correct peak spacing in addition to showing unphysical spectral negativity. 

\subsection{Example III: Laser driven dynamics of H$_2$}\label{example_driven_H2}
The present formalism is not restricted to just perturbative fields and can deal with any arbitrary external field. 
Going beyond the linear response regime, we investigate the effect of strong driving by a few-cycle ultra-fast laser pulse for this same H$_2$ model system. The system is first prepared in the ground state using the imaginary time sta-ICWF, and then the field-driven dynamics is generated by applying an electric field of the form
$E(t) = E_0\Omega(t)\sin(\omega t)$,    
with $E_0 = 0.005$a.u. and an envelope $\Omega(t)$ with a duration of 20 optical cycles. The carrier wave frequency $\omega = 0.403$ is tuned to the vertical excitation energy between the ground and second excited BOPESs at the mean nuclear position of the ground state wavefunction. A thorough description of the numerical procedure as well as the convergence behaviour of the sta-ICWF method for this model can be found in Appendix~\ref{SI_LH}.
\begin{figure}
 \includegraphics[width=\columnwidth]{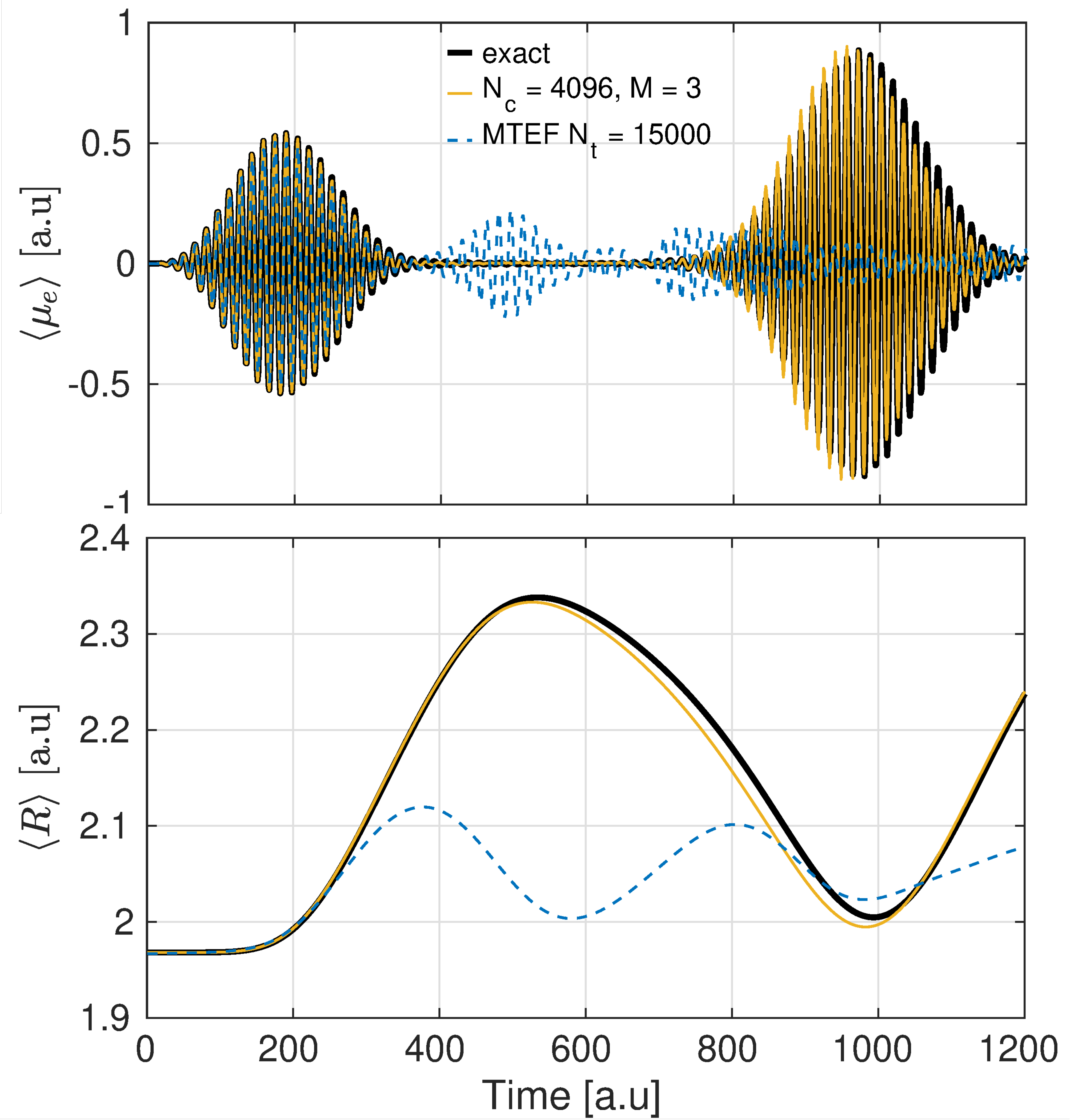}
 \caption{Top panel: Evolution of the expectation value of the dipole operator $\langle\mu_e\rangle$ for the 1D H$_2$ model system for $\text{N}_c = 4096$ (from bottom-up) and $\text{M} = 3$. Bottom panel: Evolution of the expectation value of the nuclear inter-separation $\langle R \rangle$ for the 1D H$_2$ model system for $\text{N}_c = 4096$ and $\text{M} = 3$.} 
  \label{H2_driven}
\end{figure}

The intense laser pulse creates a coherent superposition of the ground and second excited BO states whereby the bond length of the molecule increases, as shown in the bottom panel of \fref{H2_driven}. The nuclear wavepacket then eventually returns to the Franck-Condon region, creating the resurgence of the electronic dipole oscillation seen in the top panel of Fig.~\ref{H2_driven}. In the MTEF mean-field description of this process the short-time limit is rather accurately captured, while the subsequent effects of the laser pulse on the nuclear dynamics and the resurgence in the dipole response are not. These results show that the sta-ICWF method is able to capture the  electronic correlations inherent to the electronic dipole moment during the initial laser driven dynamics, as well as the electron-nuclear correlations that arise during the subsequent nonequilibrium dynamics. For this particular problem we found that $(\text{N}_c,\text{M})=(4096,3)$ was sufficient to obtain highly accurate results for both the expectation value of the electronic dipole moment (top panel of \fref{H2_driven}) and the expectation value of the inter-nuclear separation (bottom panel of \fref{H2_driven}). Further details can be found in Appendix~\ref{SI_LH}. 

\section{Time-dependent conditional wavefunctions}\label{TDCWF}
While the sta-ICWF method shows promising performance in the examples studied thus far, it faces the same limitations as any method that relies on a static basis. Perhaps the most significant aspect can be framed in terms of capturing the full support of the 
time-dependent wavefunction, which is exacerbated in cases where the time dependent state strays far from the span of the static basis. 
One strategy to address these scenarios would be to incorporate time-dependent conditional wavefunctions 
in the ICWF ansatz. Hence, we take advantage of the time-dependent version of the CWF framework introduced 
in Refs.~\cite{Albareda2014}, which relies on decomposing the exact many-body wavefunction, $\Psi(\bs x,t)$, 
in terms of time-dependent single-particle CWFs of either the electronic or nuclear subsystems as:
\begin{align}\label{CWF1}
	& \psi_{i}^{\alpha}(\bs x_i,t) :=
 \int d\bar{\bs x}_i \delta(\bar{\bs x}_i^{\alpha}(t) - \bar{\bs x}_i)  \Psi(\bs x,t),
\end{align}

Evaluating the time-dependent Schr\"odinger equation in \eref{TDSE} at $\bs x_i^{\alpha}(t)$, one can show that the CWFs in \eref{CWF1} obey the following equations of motion: \begin{align}\label{CWF_e}
    & i\frac{d}{dt} \psi_{i}^{\alpha}(t) = \left[\hat{T}_i + W^\alpha_i(t) + \eta_{i}^{\alpha}(t)\right] \psi_{i}^{\alpha}(t),
\end{align}
where $W^{\alpha}_i(\bs x_i,t) = W(\bs x_i,\bar{\bs x}_i^\alpha(t),t)$, and we remind that $W(\bs x)$ is the full electron-nuclear interaction potentials that appears in the Hamiltonian of \eref{Hamiltonian}.  In ~\eref{CWF_e}, $\eta_{i}^{\alpha}(\bs x_i,t)$ are time-dependent complex potentials containing kinetic correlations and advective terms, i.e.:
\begin{equation}\label{CP1_t}
 \eta_{i}^{\alpha}(\bs x_i,t) = \sum_{j\neq i}^{n\times N} \left(\frac{\hat T_j \Psi(t)}{\Psi(t)} \bigg|_{\bar{\bs x}_i^{\alpha}} + 
 \dot{\bs x}_j^\alpha(t) \cdot \frac{\nabla_j \Psi(t)}{\Psi(t)} \bigg|_{\bar{\bs x}_i^{\alpha}}\right) 
\end{equation}


As in the time-independent CWF framework, the conditional wavefunctions in \eref{CWF1}, represent slices of the full wavefunction taken along single-particle degrees of freedom of the two disjoint subsets. 
Each individual CWF constitutes an open quantum system, whose time-evolution is non-unitary, due to the complex potentials $\eta_{i}^{\alpha}(\bs x_i,t)$, which now include advective terms due to the inherent motion of the trajectories $\bs x^\alpha (t)$, which evolve according to Bohmian (conditional) velocity fields~\cite{Albareda2014}:
\begin{align}\label{v1}
     & \dot{\bs x}_{i}^{\alpha}(t) = \frac{1}{m_i}\text{Im} \left[ \frac{\nabla_{i} \psi_{i}^{\alpha}(\bs x_i,t)}{\psi_{i}^{\alpha}(\bs x_i,t)} \right]_{\bs x_i^\alpha(t)}. 
\end{align}
An exact solution to \eref{TDSE} can be then constructed provided we use a sufficiently large number of slices $\{\bs x^{\alpha}(t)\}$ that explore the full support of $|\Psi^\gamma(\bs x,t)|^2$ (in analogy with \fref{CWFscheme_ti}.b), i.e.:
\begin{equation}\label{reconstruction}
    \Psi(\bs x,t) = \mathcal{D}_{\bs x_i}(\psi_{i}^{\alpha}(\bs x_i,t)),
\end{equation}
where the transformations can be found in Appendix~\ref{sec:reconst}.
The one-body equations of motion in \eref{CWF_e} can be then understood each as a coupled set of non-unitary and nonlinear time-dependent problems. 

The derivation of the exact time-dependent CWF mathematical framework corresponds to the transformation of the many-body time-dependent Schr\"odinger equation to the partially co-moving frame in which all coordinates except the $i$-th move attached to the electronic and nuclear flows and only the $i$-th coordinate is kept in the original inertial frame.
Within the new coordinates, the convective motion of all degrees of freedom except for the $i$-th coordinate is described by a set of trajectories of infinitesimal fluid elements (Lagrangian trajectories), while the motion of the $i$-th degree of freedom is determined by the evolution of the CWFs in an Eulerian frame~\cite{Albareda2016}. 
The purpose of this partial time-dependent coordinate transformation is to propagate all trajectories along with the corresponding probability density flow such that they remain localised where the full molecular wavefunction has a significant amplitude.

\subsection{Time-dependent Hermitian approximation}\label{td_hermitian_sec}

In general the effective potentials in \eref{CP1_t} exhibit discontinuous steps which could introduce instabilities in a trajectory-based solution of the many-body dynamics based on \eref{CWF_e}. Therefore, in a similar manner to the time-independent case, an approximate solution can be formulated by expanding the kinetic and advective correlation potentials around the conditional coordinates $\bs x^{\alpha}(t)$, such that
\begin{align}\label{hermitian_td_approx}
    &\eta_{i}^{\alpha} (\bs x_i,t) = f(\bar{\bs x}_i^{\alpha}(t)).
\end{align}
In this limit, the kinetic and advective correlation potentials only engender a global phase that can be omitted, as expectation values are invariant under such global phase transformations. The resulting propagation scheme is restored to a Hermitian form. That is, ~\eref{CWF_e} is approximated as:
\begin{align}\label{h_e_herm}
    & i\frac{d}{dt} \psi_{i}^{\alpha}(t) = \left(\hat{T}_i + W^\alpha_i(t) \right) \psi_{i}^{\alpha}(t),
\end{align} while the trajectories ${\bs x}^\alpha(t)$ are constructed according to ~\eref{v1}.

This approximation to the time-dependent CWF formalism is clearly a major simplification of the full problem, as it recasts the  many-body time-dependent Schr\"odinger equation as a set of independent single-particle equations of motion. Despite the crudeness of the approximation in \eref{hermitian_td_approx}, the set of equations of motion in \eref{h_e_herm} has found numerous applications, e.g., in the description of adiabatic quantum molecular dynamics~\cite{Albareda2015} and quantum electron transport~\cite{Albareda2013,PhysRevB.79.075315,PhysRevB.82.085301,Albareda.ch7,oriols2007quantum}.

\section{Simulating far-from-equilibrium dynamics with Conditional wavefunctions}\label{nvar_TDSE}

In general circumstances where the kinetic and advective correlation potentials are important, we can make use of the simple Hermitian form of the conditional equations of motion in \eref{h_e_herm} to design an efficient many-body wavefunction propagator. For that, we expand the full electron-nuclear wavefunction using the ansatz: 
\begin{equation}\label{td_ansatz}
    \Psi(\mathbf{r},\mathbf{R},t) = \sum_{\alpha=1}^{\text{N}_c\text{M}} C_{\alpha}(t)\prod_{i=1}^{n\times N}\psi_{i}^{\alpha}(\bs x_i,t),
\end{equation}
where the coefficients $C_\alpha(t)$ and the CWFs $\psi_{i}^{\alpha}(\bs x_i,t)$ are initialized using the sta-ICWF method and propagated afterwards using the approximated equations of motion in ~\eref{h_e_herm} along with trajectories obeying ~\eref{v1}.

The time evolution of the coefficients $\mathbf{C}(t)$ can be then obtained by inserting the ansatz of \eref{td_ansatz} into \eref{TDSE}, 
\begin{equation}\label{coeff}
	\frac{d\mathbf{C}(t)}{dt} = -i\mathbb{S}^{-1}(t)\left(\mathbb{H}(t) - \sum_{i=1}^{n\times N}\mathbb{H}_i(t)\right) \mathbf{C}(t),
\end{equation} 
where the matrix elements of $\mathbb{S}$, $\mathbb{H}$, are defined as in Eqs. (\ref{eq5:4}) and (\ref{eq5:4_bis}), with the time dependence coming from external fields in the hamiltonian and the time dependent CWFs, while $\mathbb{H}_i$ are:
\begin{equation}\label{H_i} 
    \mathrm{H}_{i,\alpha\beta}(t) = \sum_{i=1}^{n\times N}\int d\bs x_i \psi_{i}^{\beta *} h_i^\alpha \psi_{i}^{\alpha}
    \prod_{j\neq i}^{n\times N} \int d\bs x_j \psi_{j}^{\beta *}\psi_{j}^{\alpha},
\end{equation}
where $h_i^\alpha(t)$ are the Hermitian Hamiltonians in ~\eref{h_e_herm} and $\hat{H}(t)$ is the full time-dependent Hamiltonian in \eref{TDSE}.

Obtaining these matrix elements is straightforward, involving a sum across
single body operators in \eref{eq5:4} and \eref{H_i}, and  all sums of two-body interactions across each degree of freedom in \eref{eq5:4_bis}. Note that any operator involving only a single species, e.g. the kinetic energy, is cancelled out, and thus the evolution of $\mathbf{C}$ is governed exclusively by matrix elements of operators which either fully (through $\mathbb{H}$) or conditionally (through $\mathbb{H}_i$ and $\mathbb{H}_A$) correlate the degrees of freedom.

Equations~\eqref{v1}, \eqref{h_e_herm}, and ~\eqref{coeff} define a set of coupled differential equations that hereafter will be referred to as the  dynamical ICWF (dyn-ICWF) method. One can then evaluate the expectation value of a generic observable $\langle \hat{O}(\bs x) \rangle$ as given in Eqs.~\eqref{equil_expectation} with dyn-ICWF by simply taking into account that $\psi_{i}^{\alpha}(t)$ are now time-dependent CWFs.

\subsection{Example IV: Impact electron ionization}\label{Scattering} 
The theoretical description of electron scattering remains challenging, as it is a highly-correlated problem that generally requires treatment beyond perturbation theory~\cite{scheer2017molecular,krausz2009attosecond}. 
We here study a model system of electron-Hydrogen scattering that can be exactly solved numerically~\cite{suzuki2017exact}. In atomic units, the Hamiltonian of this one-dimensional two-electron model system reads:
\begin{equation}
    \hat{H}(r_1,r_2) = \sum_{i=1}^2 \left( -\frac{1}{2}\frac{\partial^2}{\partial r_i^2} + v_{\text{ext}}(r_i) \right) + W(r_1 - r_2),
\end{equation}
where
\begin{align}
    W(r_1 - r_2) = \frac{1}{\sqrt{(r_1-r_2)^2 + 1}},
    \\
    v_{\text{ext}}(r) = \frac{1}{\sqrt{(r - 10)^2 + 1}}.
\end{align}
are respectively the soft-Coulomb interaction and the external potential that models the H atom located at $r = 10$a.u. 
The initial interacting wavefunction is taken to be a spin
singlet, with a spatial part 
\begin{equation}\label{initial_scatt}
    \Psi_0(r_1,r_2) = \frac{1}{\sqrt{2}}\left( \phi_H(r_1)\phi_{WP}(r_2) + \phi_{WP}(r_1)\phi_H(r_2) \right),
\end{equation}
where $\phi_H(r)$ is the ground state hydrogen wavefunction.
$\phi_{WP}(r)$ is an incident Gaussian wavepacket,
\begin{equation}
    \phi_{WP}(r) = \left(\frac{2\alpha}{\pi}\right)^{\frac{1}{4}} e^{\left[ -\alpha(r-r_0)^2 + ip(r-r_0) \right]},
\end{equation}
with $\alpha=0.1$, representing an electron at $r=-10$a.u.,
approaching the target atom with a momentum $p$.
\begin{figure}
 \centering
  \includegraphics[width=\linewidth]{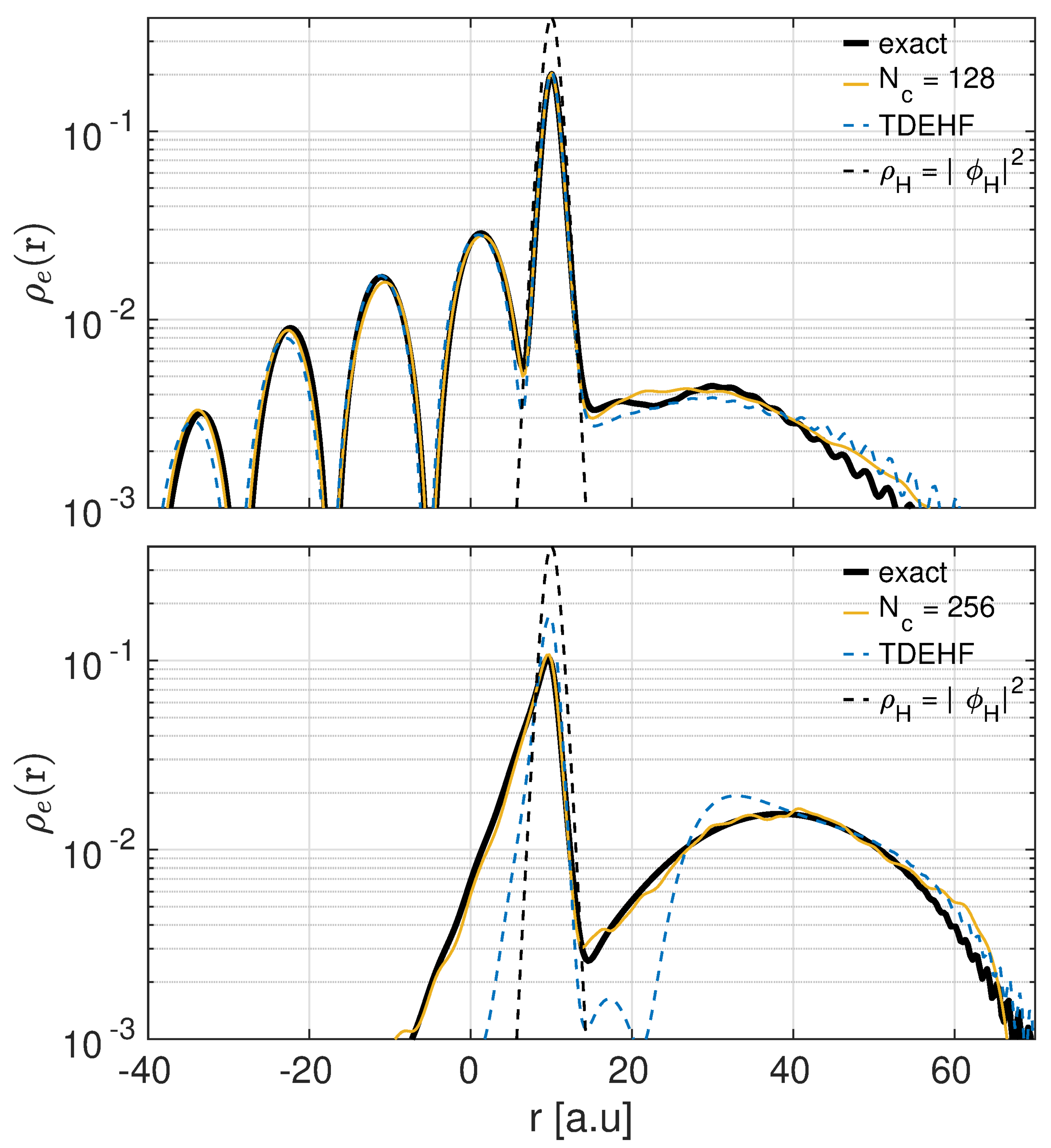}
 \caption{Top panel: reduced electron density at $t=1.8$fs for $p=0.3$a.u and $\text{N}_c = 128$. Bottom panel: reduced electron density at $t=0.85$fs for $p=1.5$a.u and $\text{N}_c = 256$ and $N_{in} = 10$}\label{rho_ee}
\end{figure}

The time-resolved picture presents scattering as a fully non-equilibrium problem, where the system starts already in a non-steady state, and so, the imaginary time sta-ICWF cannot be applied here to prepare the initial wavefunction. Instead, we stochastically sample the initial probability density $|\Psi_0(r_1,r_2)|^2$ with $\text{N}_c$ trajectories $\{r_1^\alpha(0),r_2^\alpha(0)\}$ that are used to construct CWFs $\phi_{1}^{\alpha}(r_1,0)$ and $\phi_{2}^{\alpha}(r_2,0)$ as defined in \eref{CWF1}.
A thorough description of the numerical procedure as well as the convergence behaviour of the dyn-ICWF method for this model can be found in Appendix~\ref{impact_SI}.

We study the dynamics of the electron-Hydrogen scattering by evaluating the time-dependent one-body density, $\rho_e(r_1,t) = 2\int |\Psi(r_1,r_2,t)|^2 dr_2$, for two different initial momenta, viz., $p=0.3$a.u and $p=1.5$a.u.
For $p=0.3$a.u., the energy is lower than the lowest excitation of the target (which is about $\omega = 0.4$a.u.) and hence the scattering process is elastic. In this regime, mean-field results (here represented by extended time-dependent Hartree-Fock calculations) and dyn-ICWF results with $\text{N}_c = 128$ results both capture the correct dynamics accurately.
In approaching the target atom with the larger momentum $p=1.5$a.u., the incident wavepacket collides inelastically with the target electron at around 0.24 fs, after which, a part of the wavepacket is transmitted while some is reflected back leaving the target partially ionized. In this regime, the mean-field method fails to describe the transmission process quantitatively, and the reflection process even qualitatively due to its inability to capture electron-electron correlation effects. This is in contrast with dyn-ICWF results, which quantitatively capture the correlated dynamics for $\text{N}_c = 256$, although a lower number of CWFs already reproduces qualitatively the dynamics (see Appendix~\ref{impact_SI}).

\subsection{Example V: Laser Driven Proton-Coupled Electron Transfer}\label{SM} 
We now show dyn-ICWF results for a prototypical photo-induced proton-coupled electron transfer reaction, using the Shin-Metiu model~\cite{Metiu}. The system comprises donor and acceptor ions which are fixed at a distance $L = 19.0a_0$, and a proton and an electron that are free to move in one dimension along the line connecting the donor-acceptor complex. Based on the parameter regime chosen, this model can give rise to a number of challenging situations where electron-nuclear correlations play a crucial role in the dynamics.
\begin{figure}
\includegraphics[width=\linewidth]{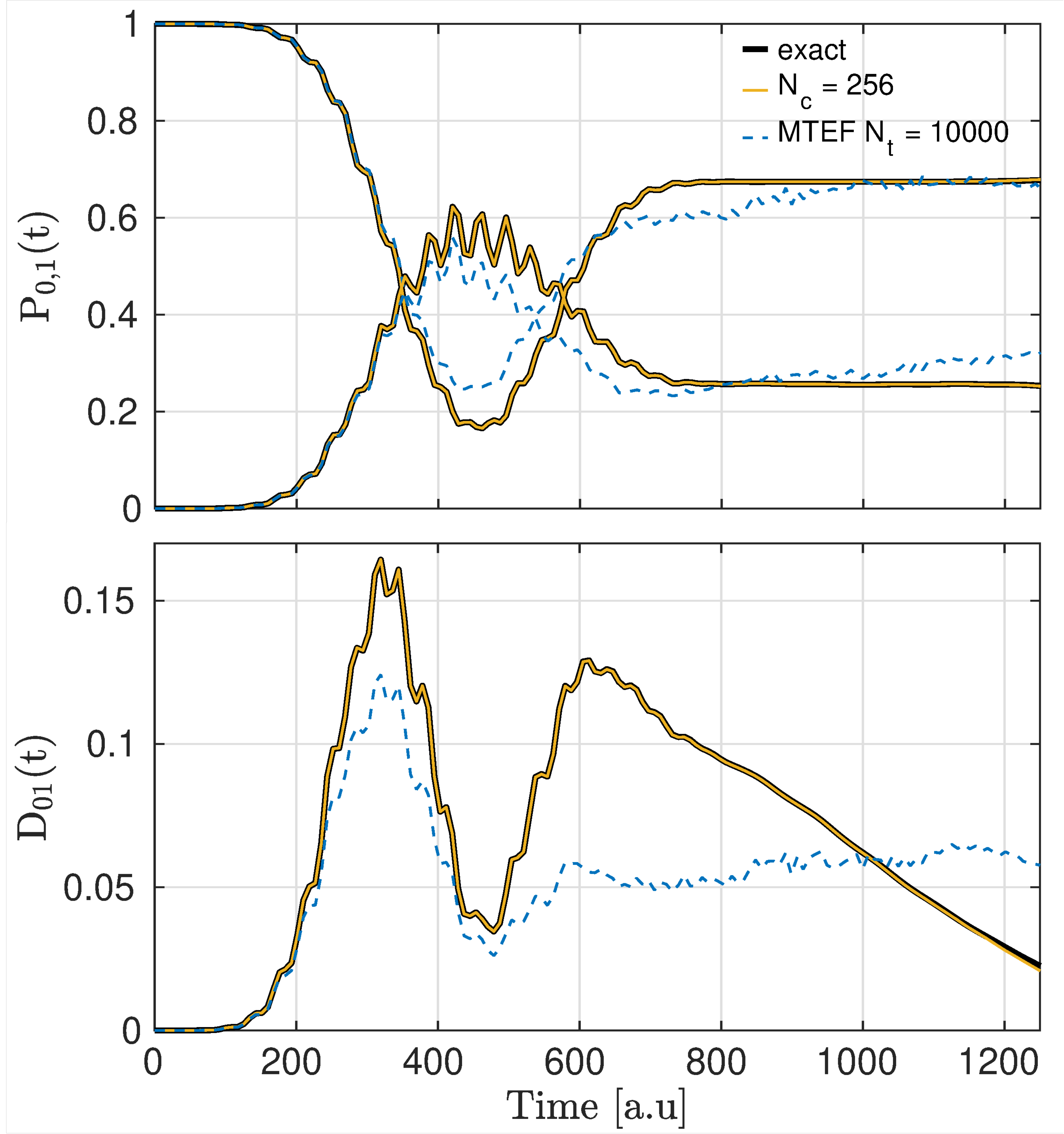}
\caption{Top panel: population dynamics of the first two adiabatic electronic states $\text{P}_{0,1}(t)$. Solid black lines correspond to exact numerical results. Solid blue and red lines correspond to dyn-ICWF results with $(\text{N}_c,\text{M}) = (256,1)$ for the ground and first excited adiabatic populations respectively. Dashed blue and red lines correspond to mean-field MTEF results. 
Bottom panel: decoherence dynamics between the ground state and first excited adiabatic electronic states, i.e., $\text{D}_{01}$. Solid black lines correspond to exact results. Solid blue line corresponds to dyn-ICWF results with $(\text{N}_c,\text{M}) = (256,1)$. ashed blue line corresponds to mean-field MTEF results.}
\label{pop_12345_SM}
\end{figure}

The total Hamiltonian for the system is,
\begin{equation}\label{metiu_ham}
 \hat H(r,R) = -\frac{1}{2m}\frac{\partial^2}{\partial r^2} - \frac{1}{2M}\frac{\partial^2}{\partial R^2} + \hat W(r,R),
\end{equation}
where $m$ is the electron mass, and $M$ is the proton mass.
The coordinates of the electron and the mobile ion are measured from the center of the two fixed ions, and are labeled $r$ and $R$, respectively. 
The full electron-nuclear potential reads:
\begin{align}\nonumber
  \hat W(r,R)  =  \frac{1}{|\frac{L}{2} - R|} + \frac{1}{|\frac{L}{2} + R|} 
- \frac{\text{erf}\big(\frac{|R - r|}{R_f}\big)}{|R - r|} \nonumber\\
- \frac{\text{erf}\big(\frac{|r - \frac{L}{2}|}{R_r}\big)}{|r - \frac{L}{2}|} 
- \frac{\text{erf}\big(\frac{|r + \frac{L}{2}|}{R_l}\big)}{|r + \frac{L}{2}|} - (r - R)E(t),
\end{align}
where $\text{erf}()$ is the error function. 
The parameter regime studied for this model ($R_f = 5a_0$, $R_l = 4a_0$ and $R_r = 3.1a_0$) and are chosen such that the ground state BOPES, $\epsilon^{(1)}_{BO}$, is strongly coupled to the first excited adiabatic state, $\epsilon^{(2)}_{BO}$, around the mean nuclear equilibrium position $R_{eq} = -2a_0$. The coupling to the rest of the BOPESs is negligible. 

We set the system to be initially in the full electron-nuclear ground state obtained from out imaginary time propagation method described above, i.e., $\Psi(r,R,0) = \Psi^0(r,R)$. We then apply an external strong electric field, $E(t) = E_0 \Omega(t) \sin(\omega t)$, with $E_0 = 0.006$a.u, $\Omega(t) = \sin(\pi t/20)^2$ and $\omega = \epsilon^{1}_{BO}(R_{eq}) - \epsilon_{BO}^{0}(R_{eq})$.
The external field induces a dynamics that involves a passage through an avoided crossing between the first two BOPESs, with further crossings occurring at later times as the system evolves. When the system passes through the nonadiabatic coupling region, the electron transfers probability between the ground state and the first excited state. This is shown in the top panel of \fref{pop_12345_SM}, where we monitor the BO electronic state populations $\text{P}_n(t)$ (whose definition can be found in Appendix~\ref{SM_SI}). 
As a result of the electronic transition, the reduced nuclear density changes shape by splitting into two parts representing influences from both ground and excited state BOPESs. This can be seen in the bottom panel of \fref{pop_12345_SM}, where, as a measure of decoherence, we use the indicator $\text{D}_{nm}(t)$ (whose definition can be found in Appendix~\ref{SM_SI}).
As nonadiabatic transitions occur the system builds up a degree of coherence which subsequently decays as the system evolves away from the coupling region. 

As shown in \fref{pop_12345_SM}, the dyn-ICWF method reaches quantitative accuracy for $(\text{N}_c,\text{M}) = (256,1)$, and vastly outperforms the multi-trajectory Ehrenfest mean-field method in describing both the adiabatic populations and the decoherence measure. More specifically, while both the dyn-ICWF method and MTEF dynamics correctly capture the exact adiabatic population dynamics at short times, the latter breaks down at long times as it fails to capture the qualitative structure of the time-evolving indicator of decoherence. Noticeably, all these aspects of this problem are qualitatively well decribed by the dyn-ICWF method using only $(\text{N}_c,\text{M}) = (16,1)$ (these results can be found in Appendix~\ref{SM_SI}).

\subsection{Example VI: Berry phase effects and molecular conical intersection}\label{Berry} 
\begin{figure}
    \centering
    \includegraphics[width=\linewidth]{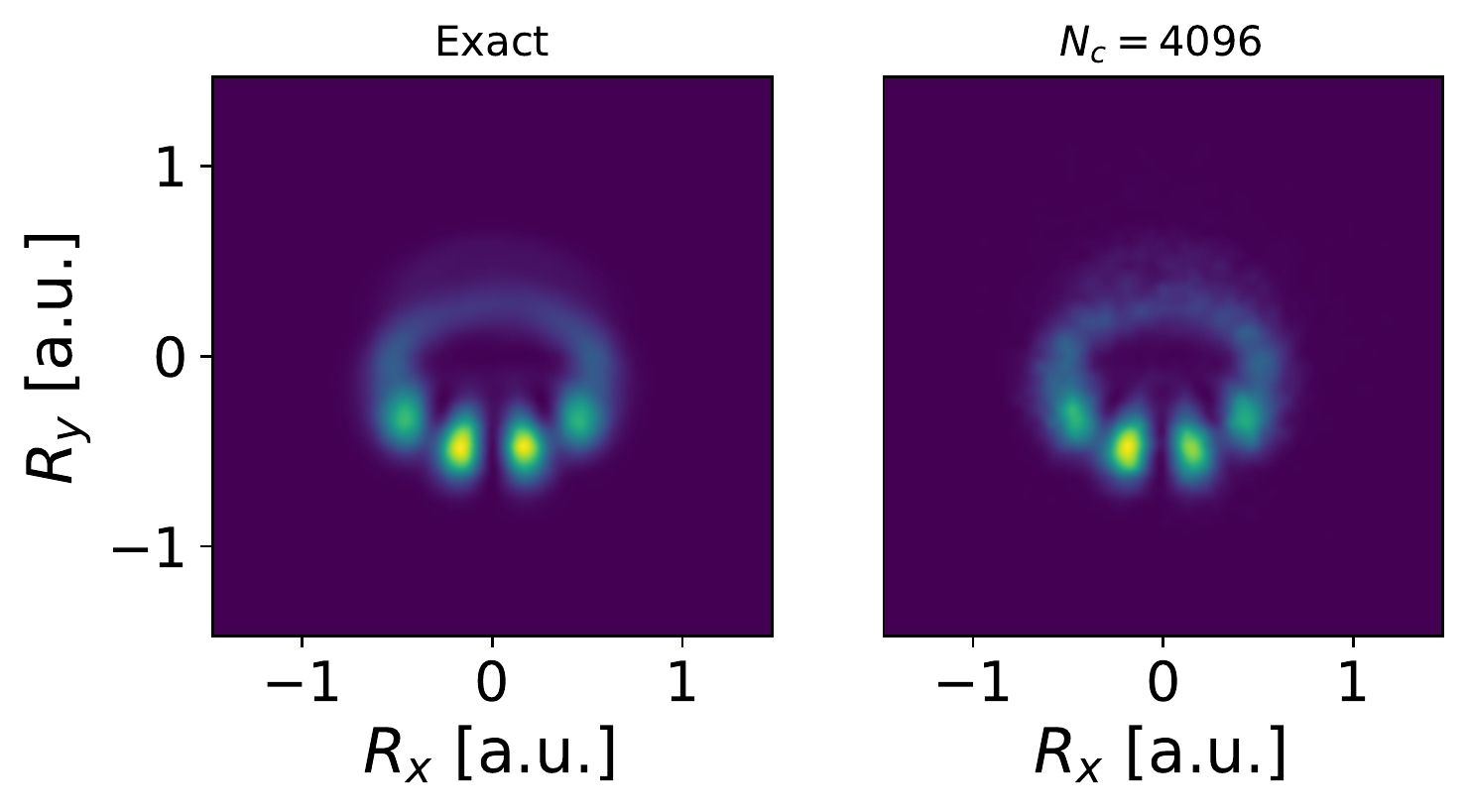}
    \caption{The exact and dyn-ICWF reduced nuclear density showing the interference pattern after having traversed the conical intersection at the origin. }
    \label{fig:4DSM}
\end{figure}
We next study dynamics around conical intersections (CIs) using a minimal generalization of the above Shin-Metiu model first proposed by Gross and co-workers~\cite{min2014molecular}, and extended further by Schaupp and Engel~\cite{schaupp2019classical}. The model consists of a quantized electron and proton that can move in two Cartesian directions, along with two fixed `classical' protons, $\mathbf{R}_1,\ \mathbf{R}_2$. A CI occurs in this model when (treating the quantized proton as a BO parameter) the protons are in a $D_{3h}$ geometry. 
The potential energy is,
\begin{align}\label{4D_SM_W}
    W(\mathbf{r,R} ) = -\frac{1}{\sqrt{a + |\bf{r}-\bf{R}|^2}} - \frac{1}{\sqrt{a + |\bf{r}-\bf{R}_1|^2}} \nonumber\\
    - \frac{1}{\sqrt{a + |\bf{r}-\bf{R}_2|^2}} + \frac{1}{\sqrt{b + |\bf{R}_1-\bf{R}_2|^2}} \nonumber\\
    + \frac{1}{\sqrt{b + |\bf{R}-\bf{R}_1|^2}} + \frac{1}{\sqrt{b + |\bf{R}-\bf{R}_2|^2}} + \left(\frac{|\bf{R}|}{R_0}\right)^4,
\end{align}
and we use the parameter values $a=0.5,\ b=10,\ R_0=1.5,\ \mathbf{R}_1=\left(-0.4\sqrt{3},1.2\right),\ \mathbf{R}_2=\left(0.4\sqrt{3},1.2\right)$. 

We initialize the total system wavefunction as a direct product of the first excited electronic BO state and a nuclear Gaussian state centered at $\mathbf{R}_0 =(0,0.4)$ with standard deviation $\sigma^2=5$. For this placement of $\mathbf{R}_1,\ \mathbf{R}_2$ the CI occurs at the origin, and in the BO picture, the initial nuclear wavepacket ``falls towards" the CI (see \fref{conical_intersection} in Appendix~\ref{Berry_SI}), while the Berry phase associated with the two possible paths around the CI cause an interference pattern to develop.
Using dyn-ICWF and \textit{propagating entirely in the real space grid picture}, this characteristic interference pattern can also be captured. Therefore, while not depending on the BO picture (beyond defining the initial state) the dyn-ICWF method retains the correct Berry curvature effects. See Appendix~\ref{Berry_SI} for further details on the dyn-ICWF calculation.

\section{Conclusions}\label{CONCL}
In this work we have introduced an exact mathematical framework that avoids the standard separation between electrons and nuclei and hence enables a unified treatment of molecular structure and nonadiabatic dynamics without relying on the construction and fit of Born-Oppenheimer potential-energy surfaces and the explicit computation of nonadiabatic couplings.

We have introduced a time-independent conditional wavefunction theory, which is an exact decomposition and recasting of the static many-body problem that yields a set of single-particle conditional eigenstates. 
Based on the imaginary time propagation of a stochastic ansatz made of approximated conditional eigenstates, the resulting method, called sta-ICWF, is able to accurately capture electron-electron correlations intrinsic to molecular structure. 
A real-time counterpart of the above method has been also derived following the Dirac-Frenkel variational procedure, and its combination with the imaginary time version yields an accurate method for solving out of equilibrium properties of molecular systems where nonadiabatic electron-nuclear correlations are important. 
This has been shown by reproducing the exact structural, linear response, and non-perturbatively driven response properties of an exactly solvable one-dimensional H$_2$ model system that standard mean-field theories fail to describe.

We have also considered a broader class of conditional wavefunctions that was formally introduced through time-dependent conditional wavefunction theory, yielding a set of coupled single-particle equations of motion. An approximated set of these time-dependent conditional wavefunctions are utilized as time-dependent basis of a stochastic wavefunction ansatz that is meant to describe observables that are relevant to far-from-equilibrium processes. The resulting propagation technique (called dyn-ICWF) in combination with sta-ICWF provides a fully self-consistent approach and, moreover, the method achieves quantitative accuracy for situations in which mean-field theory drastically fails to capture qualitative aspects of the combined electron-nuclear dynamics.

Importantly, the conditional decomposition holds for an arbitrary number of subsets (up to the total number of degrees of freedom in the system), and applies to both fermionic and bosonic many-body interacting systems. Our developments thus provide a general framework to approach the many-body problem in and out of equilibrium for a large variety of contexts. For example, using conditional wavefunctions in a form compatible with time-dependent density functional theory, in connection with alternative tensor network decompositions, or in combination with classical/semiclassical limits for specified degrees of freedom, are particularly appealing routes to follow, and work in this direction is already in progress~\cite{OCTOPUS}. 
Furthermore, the extension to periodic systems is currently under investigation and should allow the \textit{ab initio} description of driven electron-lattice dynamics such as for example laser driven heating and thermalisation~\cite{milder2021measurements,bonitz2020ab,PhysRevMaterials.3.115203,wang2015time,chen2020exciton,bernardi2014ab}, correlated lattice dynamics~\cite{eichberger2010snapshots,lan2019ultrafast,konstantinova2018nonequilibrium} and phase transitions~\cite{cudazzo2008ab,gartner2020signatures,shin2021quantum}.



\begin{acknowledgements}
This work was supported by the European Research Council (ERC-2015-AdG694097), the Cluster of Excellence 'CUI: Advanced Imaging of Matter' of the Deutsche Forschungsgemeinschaft (DFG) - EXC 2056 - project ID 390715994, Grupos Consolidados (IT1249 - 19), and the SFB925 “Light induced dynamics and control of correlated quantum systems". The Flatiron Institute is a division of the Simons Foundation.
We also acknowledge financial support from the  JSPS KAKENHI Grant Number 20K14382, the Spanish Ministerio de Econom\'{i}a y Competitividad, Projects No. PID2019-109518GB-I00, CTQ2017-87773-P/AEI/ FEDER, Spanish Structures of Excellence Mar\'{i}a de Maeztu program through grant MDM-2017-0767 and Generalitat de Catalunya, Project No. 2017 SGR 348.
\end{acknowledgements}


\bibliography{bibliography}

\appendix
\counterwithin{figure}{section}


\section{Definition of the ``reassembling'' transformation  $\mathcal{D}_{\bs x_i}$ of \eref{reconstruction_ti}}\label{sec:reconst}

Here, we consider a reconstruction of the full wavefunction $\Psi^\gamma(\bs x)$ from conditional wavefunctions defined as in \eref{CES1} of the main text, i.e.:
\begin{equation}
	\psi_{i}^{\alpha,\gamma}(\bs x_i) :=
 \int d\bar{\bs x}_i \delta(\bar{\bs x}_i^{\alpha} - \bar{\bs x}_i) \Psi^\gamma(\bs x),	\label{eq:cond-wf-elec}
\end{equation}
Here the index $\alpha \in \{1, 2, \dots, \text{N}_c\}$ denotes the particular conditional slice, and $\bar{\bs x}_i = (\bs x_1,...,\bs x_{i-1},\bs x_{i+1},...,\bs x_{n\times N})$ are the coordinates of all degrees of the system except $\bs x_i$. 
Similarly, $\bar{\bs x}_i^{\alpha} = (\bs x_1^{\alpha},...,\bs x_{i-1}^{\alpha},\bs x_{i+1}^{\alpha},...,\bs x_{n\times N}^{\alpha})$ are the position of all system's degrees of freedom except $\bs x_i$.

Assuming that the conditional sampling points, $\bar{\bs x}_i^{\alpha}$ are distributed according to a normalized distribution $\mathcal{N}(\bar{\bs x}_i^{\alpha})$, one can approximately reconstruct the full wavefunction based on the interpolation with a Gaussian function $G^{\sigma}(\bar{\bs x}_i)$ with a given width $\sigma$ as:
\begin{equation}
\Psi^{Rec,\gamma}_{\text{N}_c, \sigma}(\bs x):= \frac{
\sum^{\text{N}_c}_{\alpha=1}\frac{1}{\mathcal{N}(\bar{\bs x}_i^{\alpha})}
G^{\sigma}(\bar{\bs x}_i-\bar{\bs x}_i^{\alpha}) \psi_{i}^{\alpha,\gamma}(\bs x_i)
}{
\sum^{\text{N}_c}_{\alpha=1}\frac{1}{\mathcal{N}(\bar{\bs x}_i^{\alpha})}
G^{\sigma}(\bar{\bs x}_i-\bar{\bs x}_i^{\alpha}).
}. \label{eq:approx-reconst}
\end{equation}
In this way, the full wavefunction is reconstructed as a Gaussian weighted average: in the numerator of~\eref{eq:approx-reconst}, the contribution from each conditional slice $\alpha$ is weighted with a Gaussian distribution, and it becomes larger if the evaluated point, $\bar{\bs x}$, is closer to the sampling point $\bar{\bs x}^\alpha$. To compensate the non-uniform sampling distribution contribution, the interpolation weight is divided by the distribution function $\mathcal{N}(\bar{\bs x}_i^{\alpha})$. In addition, the denominator of~\eref{eq:approx-reconst} ensures normalization of the interpolation weight.

By considering a dense sampling ($\text{N}_c \rightarrow \infty$), the reconstructed wavefunction of~\eref{eq:approx-reconst} can be rewritten as:
\begin{align}
\lim_{\text{N}_c \rightarrow \infty}\Psi^{Rec,\gamma}_{\text{N}_c, \sigma}(\bs x) =
 \int d\bar{\bs x}^\alpha_i  
 G^{\sigma}(\bar{\bs x}-\bar{\bs x}_i^{\alpha}) \psi_{i}^{\alpha,\gamma}(\bs x_i),
 \label{eq:approx-reconst-02}
\end{align}
and substituting~\eref{eq:cond-wf-elec} into~\eref{eq:approx-reconst-02} one obtains:
\begin{align}\label{eq:approx-reconst-03}
\lim_{\text{N}_c \rightarrow \infty}\Psi^{Rec,\gamma}_{\text{N}_c, \sigma}(\bs x) =
 \int d\bar{\bs x}'_i  
 G^{\sigma}(\bar{\bs x}_i-\bar{\bs x}'_i)
 \Psi(\bar{\bs x}'),
\end{align}
where $\bar{\bs x}'=(\bs x'_1,\dots,\bs x'_{i-1},\bs x_i,\bs x'_{i+1},\dots, \bs x'_{n\times N}
)$. Therefore, for a dense sampling, $\Psi^{Rec,\gamma}_{\text{N}_c, \sigma}(\bs x)$ can be understood as the convolution of the full wavefunction $\Psi(\bs x)$ and the Gaussian weight $G^{\sigma}(\bar{\bs x}_i)$. 
Furthermore, in the narrow Gaussian width limit ($\sigma\rightarrow 0$), $G^{\sigma}(\bar{\bs x}_i)$ can be treated as a Dirac delta function and hence \eref{eq:approx-reconst-03} can be written as: 
\begin{equation}
\lim_{\substack{\sigma \rightarrow 0\\ \text{N}_c \rightarrow \infty}}\Psi^{Rec,\gamma}_{\text{N}_c, \sigma}(\bs x) =
\Psi(\bs x).
\end{equation}

In conclusion, one can exactly reconstruct the full electron-nuclear wavefunction in terms of conditional wavefunctions using the reassembling operator $\mathcal{D}_{\bs x_i}$ defined as:
\begin{align}
 \mathcal{D}_{\bs x_i} \left(\psi_{i}^{\alpha,\gamma}\right)\equiv
 \lim_{\substack{\sigma \rightarrow 0\\ \text{N}_c \rightarrow \infty}}
 \frac{
\sum^{\text{N}_c}_{\alpha=1}\frac{1}{\mathcal{N}(\bar{\bs x}_i^{\alpha})}
G^{\sigma}(\bar{\bs x}_i-\bar{\bs x}_i^{\alpha}) \psi_{i}^{\alpha,\gamma}(\bs x_i)
}{
\sum^{\text{N}_c}_{\alpha=1}\frac{1}{\mathcal{N}(\bar{\bs x}_i^{\alpha})}
G^{\sigma}(\bar{\bs x}_i-\bar{\bs x}_i^{\alpha})
}.
\end{align}

\section{Convergence of the real and imaginary time versions of the sta-ICWF method}\label{conv_S-ICWF}

In this section we discuss the convergence of the imaginary- and real-time sta-ICWF method for the examples in Secs.~\ref{example_BO}, \ref{example_abs}, and \ref{example_driven_H2}. For that we first notice that, due to the stochastic nature of the sta-ICWF method, given a set of sampling points $\text{N}_c$ and their conditional eigenstates $\text{M}$, we may also consider a number $\text{N}_{in}$ of different sets of $\text{N}_c$ sampled points and their associated $\text{M}$ conditional eigenstates. This can be accounted for by rewriting the expectation value of \eref{equil_expectation} as: \begin{equation}\label{equil_expectation_incoh}
    \langle \bar{O}(t)\rangle = \frac{1}{\text{N}_{in}}\sum_{p=1}^{\text{N}_{in}}\langle \hat{O}(t)\rangle_{p}.
\end{equation}
The dispersion of $\langle \bar{O}(t)\rangle$ with respect to $\text{N}_{in}$ is then quantified through its standard deviation, i.e.:
\begin{equation}\label{stdv}
    \Delta \bar{O}(t) = \sqrt{ \langle \bar{O}^2(t)\rangle - \langle \bar{O}(t)\rangle^2 }.
\end{equation}

\subsection{Ground and Excited BOPESs of H$_2$}\label{SI_BOPES}

We discuss here the convergence of the imaginary time version of the sta-ICWF method in capturing the ground state and excited state BOPESs for the H$_2$ model system introduced in \sref{example_BO}. 
Finding the BOPESs for this particular model is equivalent to solving \eref{eq:BO2} using the imaginary time evolution technique:
\begin{equation}
   \frac{d}{d\tau}\Phi^\zeta(r_1,r_2; R,\tau) = - \hat{\mathcal{H}}_{el}^\zeta \Phi^\zeta(r_1,r_2; R,\tau),
\label{imag_time_H2}    
\end{equation}
where $\left\{\Phi^\gamma(r_1,r_2; R)\right\}$ are the (complete, orthonormal) set of BO electronic states and we have defined $\hat{H}_{el}^\zeta$ as
\begin{equation}\label{H_el}    
    \hat{H}_{el}^\zeta(r_1,r_2; R) = \left( \mathbb{I} - \sum_{\xi=1}^{\zeta-1} \hat{P}^\xi \right) \hat{\mathcal{H}}_{el} \left( \mathbb{I} - \sum_{\xi=1}^{\zeta-1} \hat{P}^\xi \right),
\end{equation}
where $\hat{P}^\xi = \Phi^\xi\Phi^{\xi\dagger}$ and $\hat{\mathcal{H}}_{el} = \hat{H}-\hat{T}_{nuc}$. 

The BO electronic states, $\Phi^\gamma(r_1,r_2; R)$, are then expanded in terms of CWFs with the following simplified version of the ansatz in Eq.~\ref{eq3:1} that is specialized to the particular case of parametric nuclear dependence: 
\begin{equation}
    \Phi^\gamma(r_1,r_2; R) = \sum_{\alpha=1}^{\text{N}_c\text{M} } C_{\alpha}^{\gamma} \phi_{1}^{\alpha}(r_1;  R)\phi_{2}^{\alpha}(r_2;  R) .
\label{BO_1}
\end{equation}

Slicing points ($r_1^{\alpha},r_2^{\alpha}$) are generated by sampling from reduced one-body electronic densities, which in this case are simply chosen to be Gaussian functions $\rho_e(r_i) = A e^{-r_i^2/10}$. The conditional eigenstates $\phi_{i}^{\alpha,\nu}(r_i; R),$ for $\nu\in\{1,\ldots,M\}$ are then evaluated on each slice using the Hermitian approximation, i.e.:
\begin{equation}\label{h_eigs_e_BO}
   \left(-\frac{\hbar^2}{2m}\nabla^2_{i} + W^{\alpha}_{i}(r_i,R) \right)\phi_{i}^{\alpha,\nu}(r_i;R) = E^\nu(R) \phi_{i}^{\alpha,\nu}(r_i;R),
\end{equation}
where $W^{\alpha}_{i}(r_i,R) = W_{ee}(r_i,\bar{r_i}^\alpha) + W_{en}(r_i,R)$.
The coefficient vector $\mathbf{C}^\gamma$ is randomly initialized and then propagated in imaginary time until the target state is reached according to \eref{eq5:3} of the main text with $\hat H$ being substituted with $\hat{\mathcal{H}}_{el}$. 

To achieve converged results, a grid $(0,9]$a.u. for the internuclear separation with 181 grid points is chosen for the nuclear degrees of freedom. For the electron coordinates, the grid covers the interval [–35,+35]a.u. with 200 grid points. The fourth-order Runge-Kutta integration method~\cite{runge1895numerische,kutta1901beitrag} was used to propagate the imaginary time sta-ICWF equations of motion with a time-step $d\tau = 0.01$a.u, and the Moore-Penrose pseudo-inversion method with a tolerance of $10^{-8}$ was used to approximate the numerical inversion of the overlap matrix in \eref{eq5:4}. 
Importantly, the matrices $\mathbb{S}$ and $\mathbb{H}$ of \eref{eq5:3} need only be constructed at the initial time, requiring only the repeated multiplication of a $\text{N}_c\times \text{M}$ vector by a $\text{N}_c^2\times \text{M}^2$ matrix for the imaginary time propagation.

\begin{figure}
\includegraphics[width=\columnwidth]{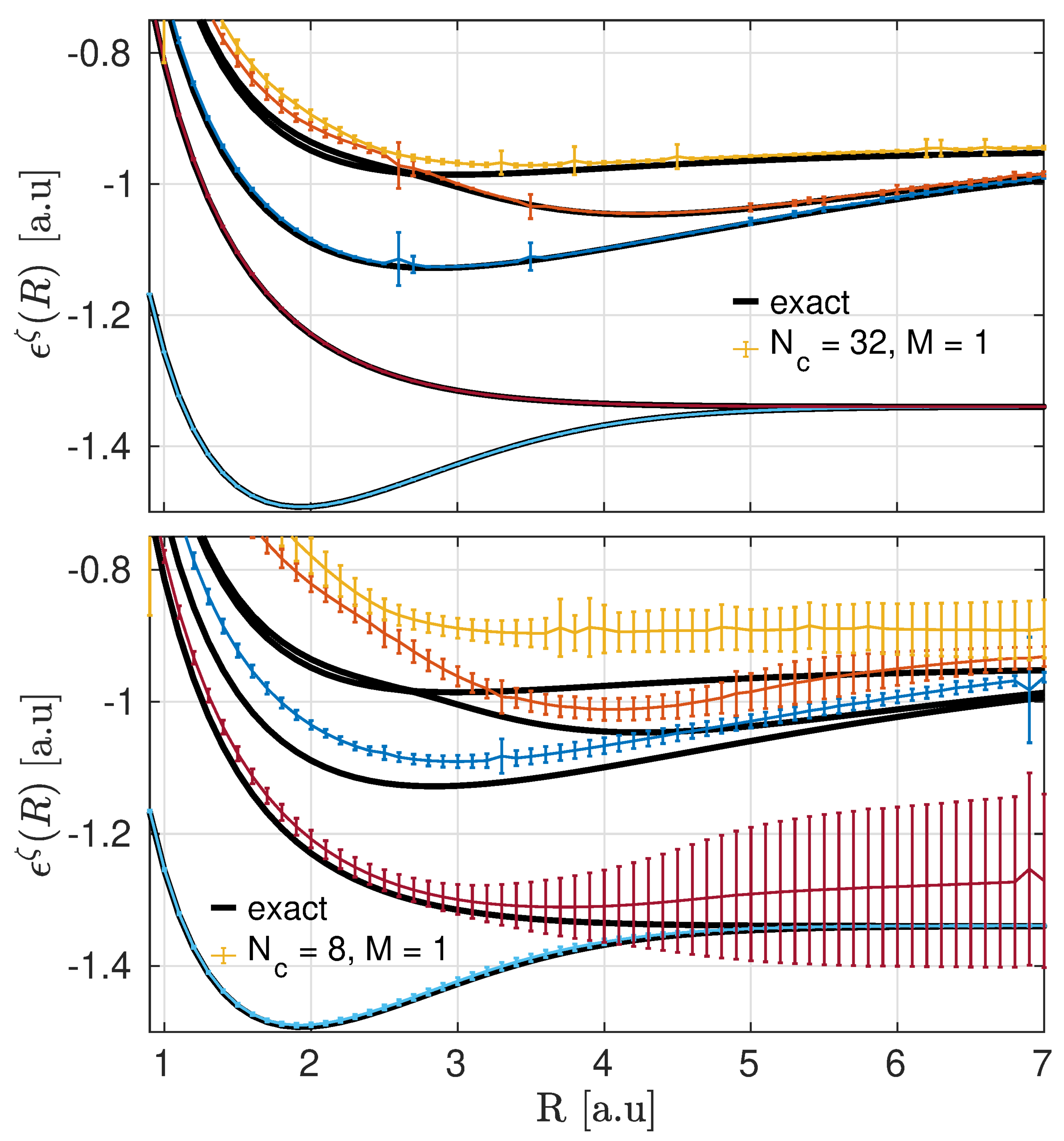}
\caption{First five BOPESs reproduced with the sta-ICWF method for $(\text{N}_c,\text{M} = (8,1)$ (bottom panel) and $(\text{N}_c,\text{M}) = (32,1)$ (top panel). These data are presented alongside with (standard deviation) error bars.}
\label{BOPES_convergence}
\end{figure}

In \fref{BOPES_convergence}, we show sta-ICWF results for the first five BOPESs for two different sets of parameters: $(\text{N}_c,\text{M})=(32,1)$ (top panel) and $(\text{N}_c,\text{M})=(8,1)$ (bottom panel).  
The sta-ICWF data is presented alongside (standard deviation) error bars defined in \eref{stdv}. 
Noticeably, even for $\text{M}=1$ (i.e., when only ground state conditional eigenstates are used in the expansion of \eref{BO_1}) the results in \fref{BOPES_convergence} demonstrate the convergence of the imaginary time sta-ICWF method to the exact BOPESs.
For large enough number of sampling points and excited CWFs, viz., $(\text{N}_c,\text{M}) \gtrsim (32,5)$, the sta-ICWF results are fully converged to the exact BOPESs and the associated error bars become negligible due to the completeness of the CWF basis.

\subsection{Ground state of H$_2$}\label{SI_GS}
We investigate here the ground-state energy for the model H$_2$ introduced in \sref{example_BO} as well as the convergence behaviour of the imaginary time version of the sta-ICWF method in capturing it. 
We aim to solve \eref{eq5:1}, which for this particular model system reduces to:
\begin{equation}
    \frac{d\Psi^{(0)}(r_1,r_2,R,\tau)}{d\tau} = - \hat{H} \Psi^{(0)}(r_1,r_2,R,\tau),
\label{eq5:1_bis}    
\end{equation}
where $\hat{H}$ is the Hamiltonian in \eref{H_H2}. For that, we choose the conditional eigenstates basis by sampling $\text{N}_c$ points ($r_1^{\alpha},r_2^{\alpha}, R^{\alpha}$) from guesses to the reduced electronic and nuclear densities $\rho_e(r_i) = A_e e^{-r_i^2/10}$ and $\rho_n(R) = A_n e^{-(R-2)^2}$ respectively. These positions are then used to construct and diagonalize the Hermitian Hamiltonians in \eref{h_eigs_e}. In this way we obtain $3\times\text{N}_c\times\text{M}$ conditional eigenstates $\left\{\phi_{1}^{\alpha,\zeta}(r_1),\phi_{2}^{\alpha,\zeta}(r_2),\chi^{\alpha,\zeta}(R)\right\}$. 

Given a random initialization of the coefficients vector $\mathbf{C}$, we then evolve it in imaginary time according to \eref{eq5:3}.
To achieve converged results, a grid $(0,9]$a.u. for the internuclear separation with 181 grid points is chosen for the nuclear degrees of freedom. For the electron coordinates, the grid covers the interval [–35,+35]a.u. with 200 grid points. The fourth-order Runge-Kutta algorithm with a tolerance of $10^{-8}$ was used to propagate the imaginary time sta-ICWF equations of motion with a time-step $d\tau = 0.01$a.u, and the Moore-Penrose pseudo-inversion method was used to approximate the numerical inversion of the overlap matrix in \eref{eq5:4}. 
Importantly, the matrices $\mathbb{S}$ and $\mathbb{H}$ of \eref{eq5:3} need only be constructed at the initial time.

From the exact symmetric ground-state wave function, we found an equilibrium separation of $\langle R \rangle = 2.2$a.u. and the ground-state energy is $E_0 = -1.4843$a.u. 
We then define the relative error of the sta-ICWF calculation with respect to the exact calculation as
$\text{E}_\text{r} = {\left | \langle \bar{H} \rangle_0 - E_0 \right |}/{\left |E_0 \right |}$, where 
\begin{equation}
    \langle \bar{H}\rangle_0 = \frac{1}{\text{N}_{in}}\sum_{n=1}^{\text{N}_{in}}\langle \Psi^{(0)} | \hat{H} | \Psi^{(0)} \rangle_{n}.
\end{equation}

\begin{figure}
\includegraphics[width=\columnwidth]{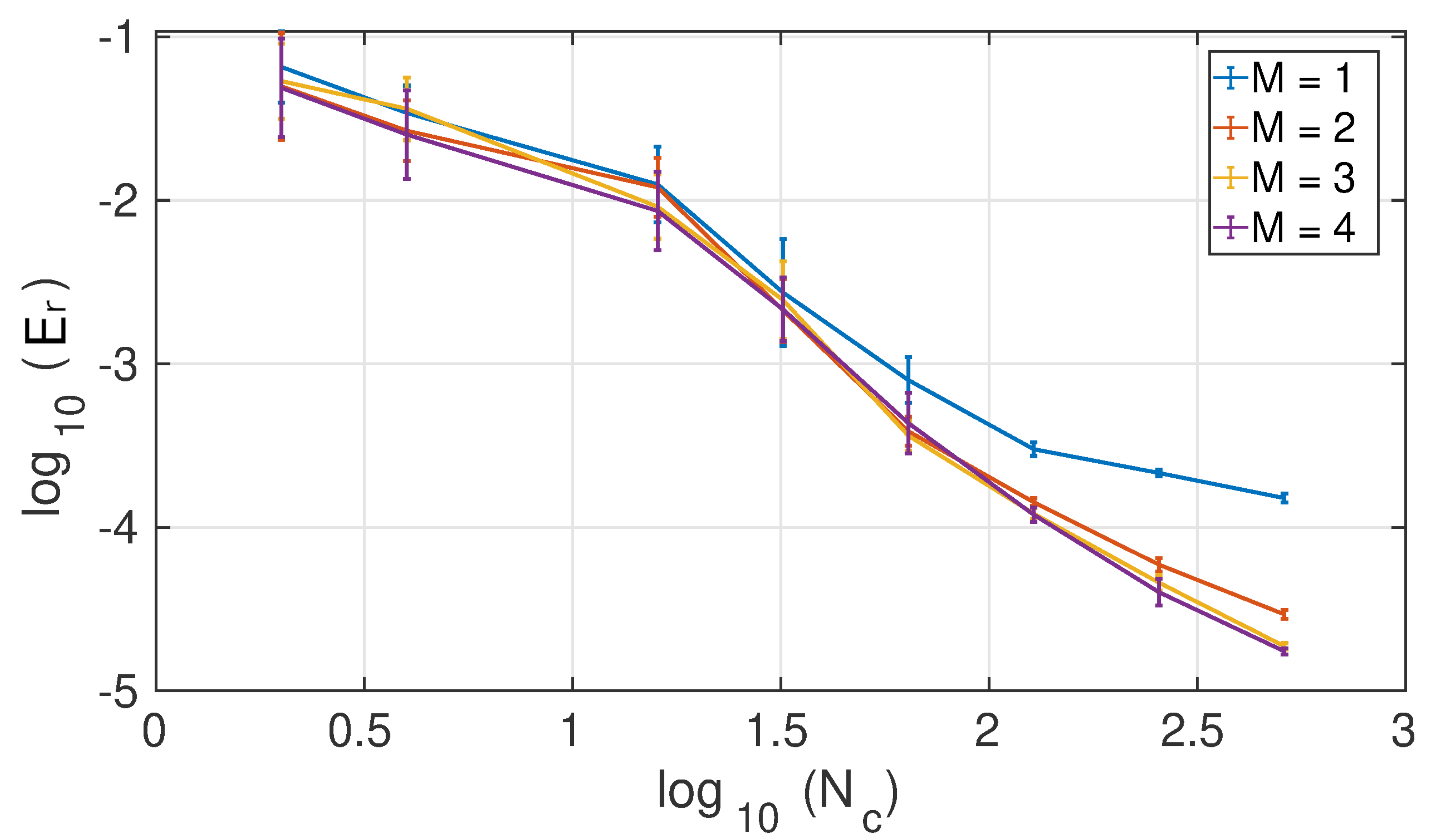}
\caption{(Left) Logarithm of the mean relative energy error $\text{E}_\text{er}$ as a function of the logarithm of the number of sampling points $\text{N}_c$ and for different number of excited CWFs $\text{M} = \{1,2,3,4\}$. Error bars represent the standard deviation of the relative error.}
\label{Energy_std}
\end{figure}

The error $\text{E}_r$ is presented in \fref{Energy_std} as a function of the number of sampling points and for different number of excited conditional eigenstates, i.e., $(\text{N}_c,\text{M})$. 
Error bars represent the standard deviation $\Delta \bar{H}_0$ defined in \eref{stdv} for a number of different initial sampling points. 
Due to the variational nature of the method, the relative error decreases with increasing number of sampling points $\text{N}_c$. 
Noticeably, even for $\text{M}=1$ (i.e., when only ground state conditional eigenstates are used in the expansion of \eref{eq3:1}) the results in \fref{Energy_std} demonstrate the convergence of the imaginary time sta-ICWF method to the exact ground state.
The convergence process is though accelerated as we allow a number of excited conditional eigenstates 
(i.e., $\text{M}> 1$) to participate in the ansatz. 
For large enough number of basis elements $\text{N}_c \times \text{M}$, the CWF bases become a complete basis of the problem. This is so independently of the initial distribution of sampling points and hence the associated error bars vanish for large enough values of $\text{N}_c \times \text{M}$.

\subsection{Optical absorption spectrum of H$_2$}\label{SI_optical}

We discuss here the convergence of the real-time version of the sta-ICWF method in capturing the optical absorption spectrum of the H$_2$ model system introduced in \sref{example_BO}.
The simulation starts with the preparation of the ground state coefficients $\mathbf{C}(0)$ using the imaginary time version of the sta-ICWF method. The relevant degree of freedom of the kick operator is then applied to each CWF, the Hamiltonian and inverse overlap matrices are reconstructed, and $\mathbf{C}$ is propagated to the desired time according to \eref{eq: ICWF prop}. A kick strength of $\kappa = 10^{-4}$a.u$^{-1}$ was sufficient to generate the kick spectra within the linear response regime and a total propagation time of $T_f=1500$a.u. was used to generate the spectra, alongside the mask function $\mathcal{M}(x=t/T_f) = 1-3x^2 + 2x^3$.

To achieve convergence, a grid $[-35,+35]$a.u. with 200 grid points is chosen for the electronic coordinates. The fourth-order Runge-Kutta algorithm was used to propagate the imaginary time sta-ICWF equations of motion with a time step $dt = 0.01$a.u, and the Moore-Penrose pseudo-inversion method with a tolerance of $10^{-8}$ was used to approximate the numerical inversion of the overlap matrix in \eref{eq5:4}. 
Again, the matrices $\mathbb{S}$ and $\mathbb{H}$ of \eref{eq5:3} need only be constructed at the initial time.

\begin{figure}
\includegraphics[width=\columnwidth]{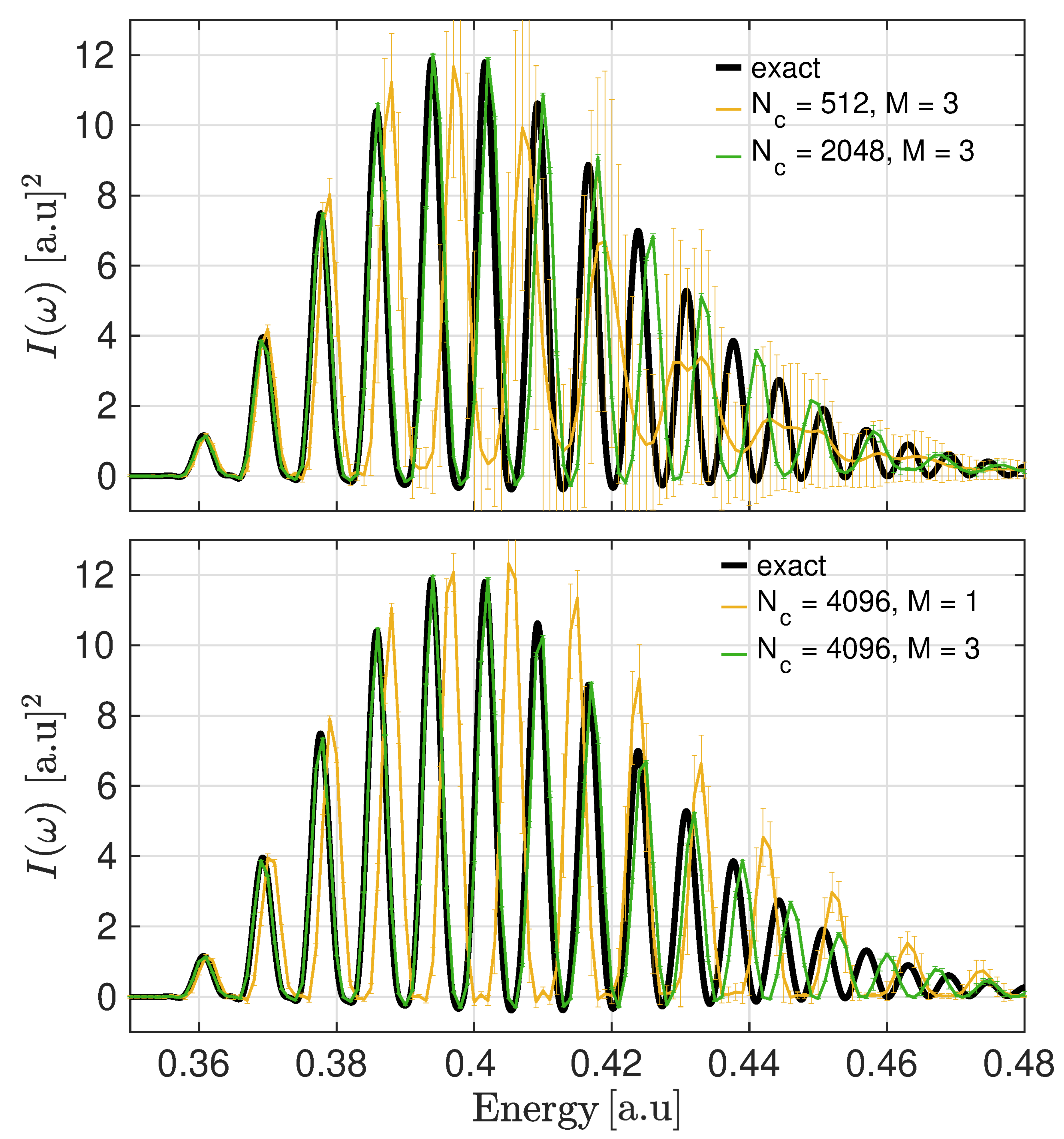}
\caption{H$_2$ spectrum for ICWF-Kick with different number of sampling points and excited CWFs. Top panel: $(\text{N}_c,\text{M}) = (512,3)$ and $(\text{N}_c,\text{M}) = (2048,3)$. Bottom panel:  $(\text{N}_c,\text{M}) = (4096,1)$ and $(\text{N}_c,\text{M}) = (4096,3)$. The results are presented alongside (standard deviation) error bars.}
\label{fig: SI-ICWF Convergence}
\end{figure}

In \fref{fig: SI-ICWF Convergence} we show convergence results for sta-ICWF calculations of the optical linear absorption spectra for four different of sets of parameters: $(\text{N}_c,\text{M}) = (512,3)$, $(\text{N}_c,\text{M}) = (2048,3)$ (top panel), and  $(\text{N}_c,\text{M}) = (4096,1)$ and $(\text{N}_c,\text{M}) = (4096,3)$ (bottom panel). 
In all of these cases, we considered a number of different initial sampling points, which have been used to calculate the associated (standard deviation) error bars as in \eref{stdv}.
As the number of conditional eigenstates basis elements in the ansatz expansion of \eref{eq3:1} increases, the variational nature of the method ensures convergence to the exact linear absorption lineshape. Similarly, the error bars shrink as the number of conditional eigenstates in the basis $\text{N}_c\times\text{M}$ allows to span the relevant part of the Hilbert space. 

\subsection{Laser driven dynamics of H$_2$}\label{SI_LH}
We discuss here the convergence of the real-time version of the sta-ICWF method in capturing the laser driven dynamics of the H$_2$ model system introduced in \sref{example_BO}. 
As explained in \sref{example_driven_H2} of the main text, the system is first prepared in the ground state using the imaginary time sta-ICWF as explained in \sref{SI_GS}, and then the field-driven dynamics is generated by applying an electric field of the form $E(t) = E_0\Omega(t)\sin(\omega t)$,    
with $E_0 = 0.005$a.u. and an envelope $\Omega(t)$ with a duration of 20 optical cycles. The carrier wave frequency $\omega = 0.403$ is tuned to the vertical excitation between the ground BO state and second excited electronic surface. 

For the dynamics we used a grid $(0,9]$a.u. for the internuclear separation with 181 grid points is chosen for the nuclear degrees of freedom. For the electron coordinates, the grid covers the interval [–35,+35]a.u. with 200 grid points. The fourth-order Runge-Kutta algorithm was used to propagate the imaginary time sta-ICWF equations of motion with a time-step $dt = 0.01$a.u, and the Moore-Penrose pseudo-inversion method with a tolerance of $10^{-8}$ was used to approximate the numerical inversion of the overlap matrix in \eref{eq5:4}. 

\begin{figure}
  \includegraphics[width=\columnwidth]{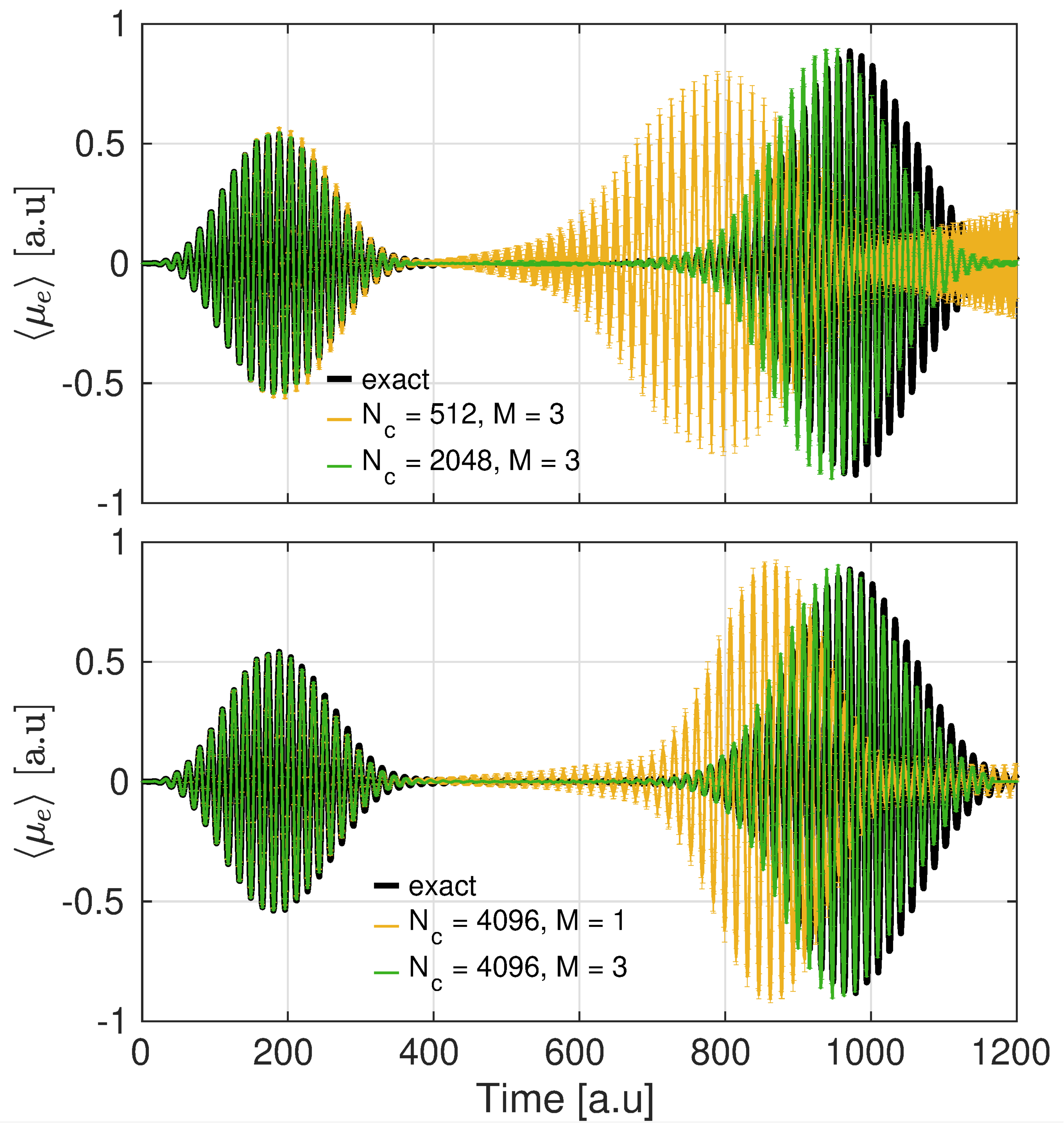}
 \caption{Evolution of the expectation value of the dipole operator $\langle\mu_e\rangle$ for the 1D H$_2$ model system for a number of conditional basis configurations. Top panel: $(\text{N}_c,\text{M}) = (512,3)$, $(\text{N}_c,\text{M}) = (4096,3)$. Bottom panel:  $(\text{N}_c,\text{M}) = (4096,1)$ and $(\text{N}_c,\text{M}) = (4096,3)$. These data is presented along with (standard deviation) error bars.}
  \label{e_dipole_H2}
\end{figure}

In Figs.~\ref{e_dipole_H2} we show convergence results for the real-time sta-ICWF calculation of the electronic dipole moment $\langle \hat{\mu}_e \rangle$. 
We considered four different sta-ICWF configurations, viz., $(\text{N}_c,\text{M}) = (512,3)$, $(\text{N}_c,\text{M}) = (4096,3)$ (in the top panel), and  $(\text{N}_c,\text{M}) = (4096,1)$ and $(\text{N}_c,\text{M}) = (4096,3)$ (in the bottom panel). 
As the number of CWFs in the ansatz expansion of \eref{eq3:1} increases, the variational nature of the method ensures convergence to the exact dynamics. 
The deviation from the exact results does grow with increasing time lapse, although this is ameliorated with increasing either $\text{N}_c$ and/or $M$, and can in principle be eliminated at large enough values of these parameters. Similarly, the error bars become negligible when the CWFs bases expand the full support of the Hilbert space explored during the dynamics. This happens for $(\text{N}_c,\text{M}) \gtrsim (4096,3)$.

\section{Convergence of the dyn-ICWF method}\label{conv_D-ICWF}
In this section we discuss the convergence behaviour of the dyn-ICWF method for the examples of Secs.~\ref{Scattering}, \ref{SM}, and \ref{Berry}. As it happened for the sta-ICWF method, the stochastic nature of the dyn-ICWF method allows us to consider a number $\text{N}_{in}$ of different initial sampling points for a given set of parameters $(\text{N}_c,\text{M})$.
This is taken into account by writing expectation values as in \eref{equil_expectation_incoh} and its standard deviation as in \eref{stdv}.

\subsection{Impact electron ionization}\label{impact_SI}
We discuss here the convergence behaviour of the dyn-ICWF method in capturing the laser driven proton-coupled electron transfer described in \sref{Scattering}.

The time-resolved picture presents scattering as a fully non-equilibrium problem, where the system starts already in a non-steady state, and so, the imaginary time sta-ICWF cannot be applied here to prepare the initial wavefunction. Instead, we stochastically sample the initial probability density $|\Psi_0(r_1,r_2)|^2$ with $\text{N}_c$ trajectories $\{r_1^\alpha(0),r_2^\alpha(0)\}$ that are used to construct CWFs $\phi_{1}^{\alpha}(r_1,0)$ and $\phi_{2}^{\alpha}(r_2,0)$ as defined in \eref{CWF1}. These CWFs are then used to construct the ansatz in \eref{td_ansatz} with an initial $\bf{C}$ vector that is obtained using $\bs{C}(0) = \mathbb{S}^{-1}\bs{G}$,
where $\bf{G}$ is the vector containing the overlap between the initial wavefunction and the CWFs, i.e.:
\begin{equation}\label{G}
G_\alpha =  \iint dr_1 dr_2 \phi_{1}^{\alpha *}(r_1,t)\phi_{2}^{\alpha *}(r_2,t) \Psi_0(r_1,r_2).
\end{equation}
Given $\bs{C}(0)$, and $\phi_{1}^{\alpha}(r_{1},0)$ and $\phi_{2}^{\alpha}(r_{2},0)$ for an ensemble of sampling points $\{r_1^\alpha(0),r_2^\alpha(0)\}$, these objects are then propagated according to the dyn-ICWF equations of motion. 

To achieve converged results, we choose the size of the simulation box to be $150\times 150$a.u$^2$ with an homogeneous grid consisting of 500 grid points in each direction. The fourth-order Runge-Kutta algorithm was used to propagate the dyn-ICWF equations of motion with a time-step $dt = 0.01$a.u, and the Moore-Penrose pseudo-inversion method with a tolerance of $10^{-8}$ was used to approximate the numerical inversion of the overlap matrix in \eref{coeff}.

\begin{figure}
  \centering
  \includegraphics[width=\linewidth]{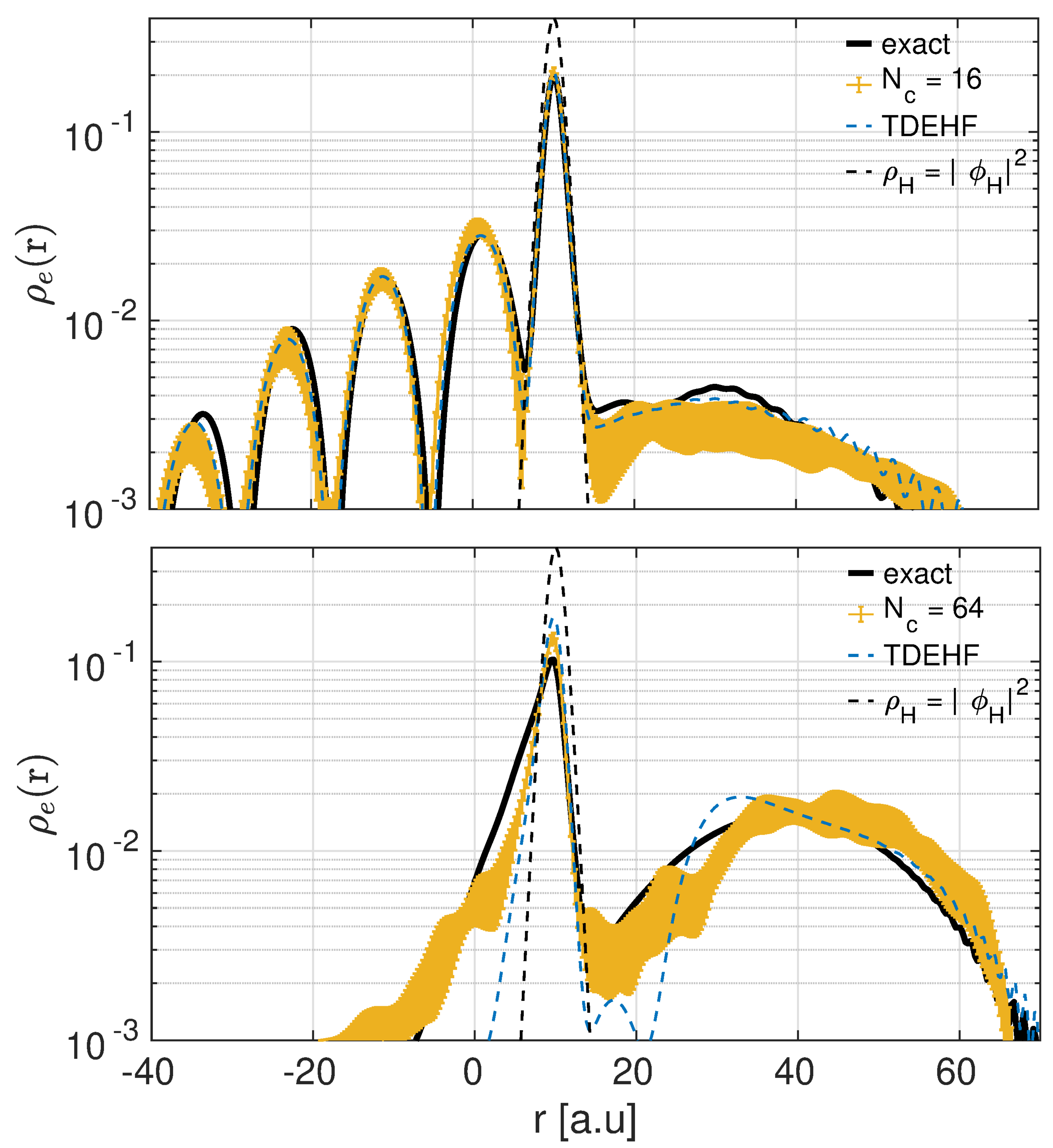}
 \caption{Top panel: reduced electron density at $t=1.8$fs for $p=0.3$a.u and $(\text{N}_c,\text{M}) = (16,1)$. Bottom panel: reduced electron density at $t=0.85$fs for $p=1.5$a.u and $(\text{N}_c,\text{M}) = (64,1)$.}\label{rho_elastic_SI}
\end{figure}

In \fref{rho_elastic_SI}, we show the one-body electronic density $\rho_e(r_1,t)$, for two different initial momenta and final times, viz., $p=0.3$a.u and $p=1.5$a.u and $t = 1.8$fs and $t = 0.85$fs.
For $p=0.3$a.u., a very small number of CWFs ($(\text{N}_c,\text{M}) = (16,1)$) is already able to capture the correct dynamics quantitatively.
In approaching the target atom with the larger momentum $p=1.5$a.u., the conventional mean-field method fails to describe the ionization process due to the lack of electron-electron correlation effects. This is in contrast with dyn-ICWF results, which qualitatively captures the correlated dynamics for a small number of CWFs $(\text{N}_c,\text{M}) = (64,1)$.

\subsection{Laser Driven Proton-Coupled Electron Transfer}\label{SM_SI}
We discuss here the convergence behaviour of the dyn-ICWF method in capturing the laser driven proton-coupled electron transfer described in \sref{SM}. 
We suppose the system to be initially seating in the full electron-nuclear ground state, i.e., $\Psi(r,R,0) = \Psi^0(r,R)$.
This state is prepared using the imaginary time version of the sta-ICWF method with ground state CWFs only (i.e., $\text{M}=1$). 
The sta-ICWF provides as output the initial expansion coefficients $\bs C(0)$ and the ground state CWFs $\phi_{i}^{\alpha}(\mathbf{r}_i,0)$ and $\chi_J^{\alpha}(\mathbf{R}_J,0)$.
We then apply an external strong electric field, defined in \sref{SM} of the main text, and the coefficients and the CWFs are propagated using the dyn-ICWF equations of motion. 

To achieve converged results, a grid $[-9,9]$a.u. with 301 grid points is chosen for the nuclear degrees of freedom. For the electron coordinates, the grid covers the interval [–75,+75]a.u. with 250 grid points. The fourth-order Runge-Kutta algorithm was used to propagate the dyn-ICWF equations of motion with a time-step $dt = 0.1$a.u, and the Moore-Penrose pseudo-inversion method with a tolerance of $10^{-8}$ was used to approximate the numerical inversion of the overlap matrix in \eref{coeff}.

By introducing the so-called Born-Huang expansion of the molecular wavefunction, $\Psi(\mathbf{r},\mathbf{R},t) = \sum_n \Phi_\mathbf{R}^{(n)}(\mathbf{r},t)\chi^{(n)}(\mathbf{R},t)$, we then monitor the dynamics through the BO electronic state populations:
\begin{equation}\label{ad_pop}
    \text{P}_n(t) = \int d\mathbf{R} |\chi^{(n)}(\mathbf{R},t)|^2,
\end{equation} 
and the overlap integral of projected nuclear densities evolving on different BOPESs:
\begin{equation}\label{decoh}
    \text{D}_{nm}(t) = \int d\mathbf{R} |\chi^{(n)}(\mathbf{R},t)|^2|\chi^{(m)}(\mathbf{R},t)|^2.
\end{equation} 
\begin{figure}
\includegraphics[width=\linewidth]{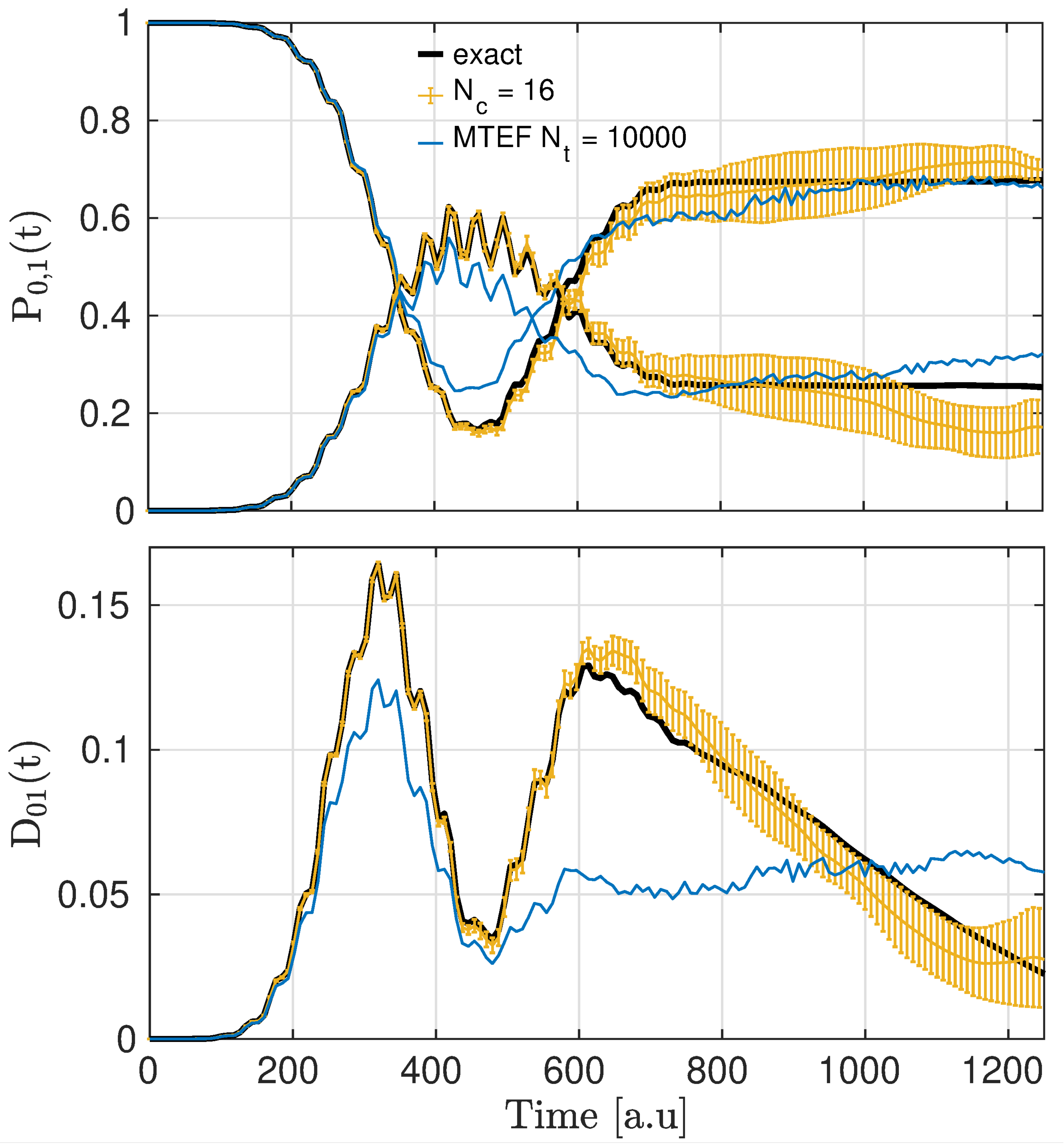}
\caption{Top panel: population dynamics of the first two adiabatic electronic states $\text{P}_{0,1}(t)$. Solid black lines correspond to exact numerical results. Solid blue and red lines correspond to dyn-ICWF results with $(\text{N}_c,\text{M}) = (16,1)$ for the ground and first excited adiabatic populations respectively. 
Bottom panel: decoherence dynamics between the ground state and first excited adiabatic electronic states, i.e., $\text{D}_{01}$. Solid black lines correspond to exact results. Solid blue line corresponds to dyn-ICWF results with $(\text{N}_c,\text{M}) = (16,1)$.}
\label{conv_pop_12}
\end{figure}

In \fref{conv_pop_12}, we show dyn-ICWF results for $(\text{N}_c,\text{M}) = (16,1)$. This very small number of CWFs, even if associated to large deviations across different stochastic particle placements, is able to captured nearly quantitatively both the adiabatic populations and the decoherence indicator of \eref{ad_pop} and \eref{decoh}. 
This results demonstrate that the dyn-ICWF technique achieves quantitative accuracy for situations in which mean-field theory drastically fails to capture
qualitative aspects of the dynamics using three orders of magnitude fewer trajectories than a mean-field simulation.

\subsection{Berry phase effects and molecular conical intersection}\label{Berry_SI}
\begin{figure}
  \includegraphics[width=\linewidth]{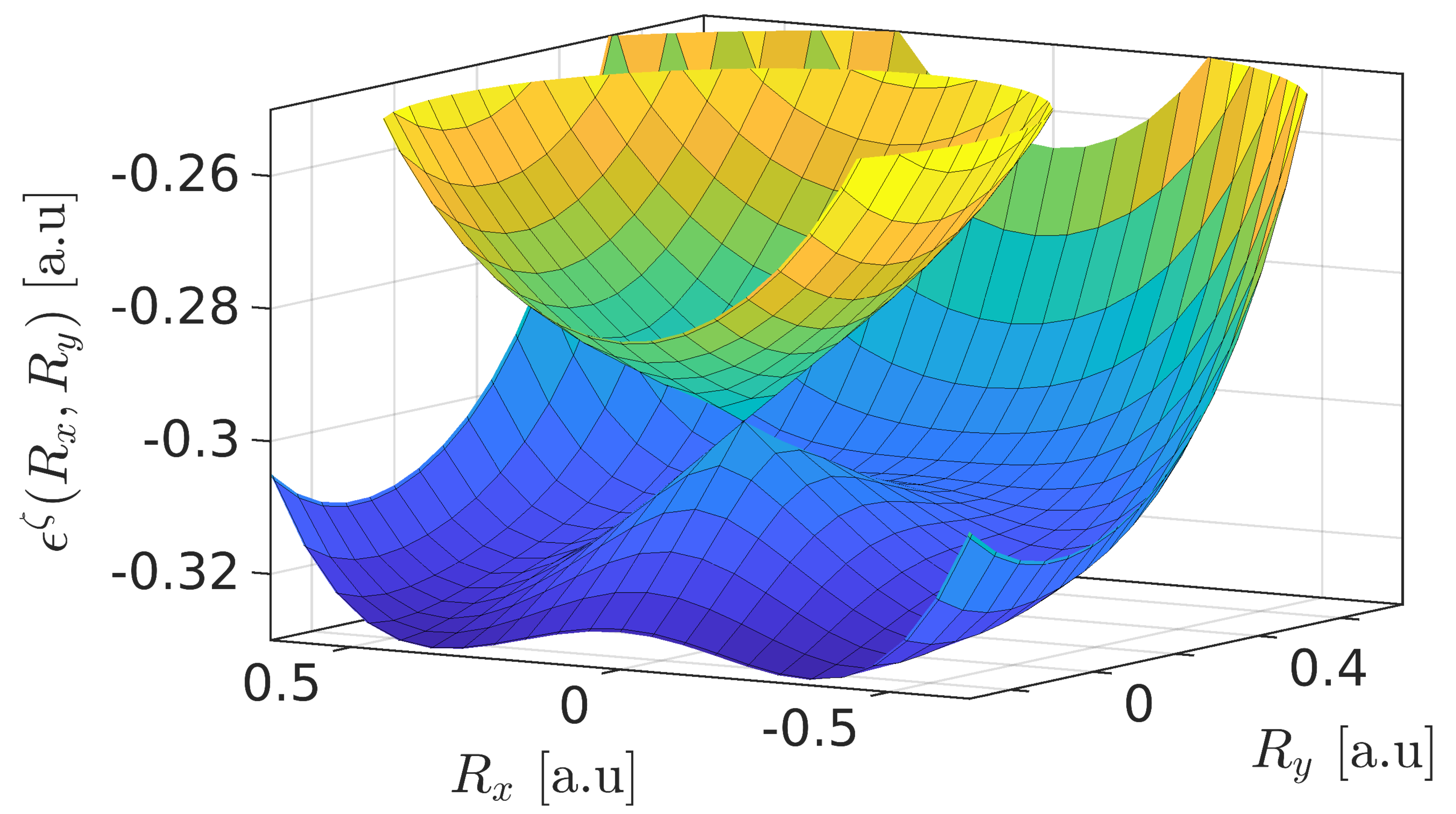}
 \caption{BOPESs for the first two excited states with electronic quantum numbers $\zeta = 1$ (lower surface) and $\zeta = 2$. As mentioned in the main text, the initial nuclear state of is intialized as a gaussian centered at $\mathbf{R}=(0,0.4)$ on the lower surface.}\label{conical_intersection}
\end{figure}
\begin{figure}
  \includegraphics[width=\linewidth]{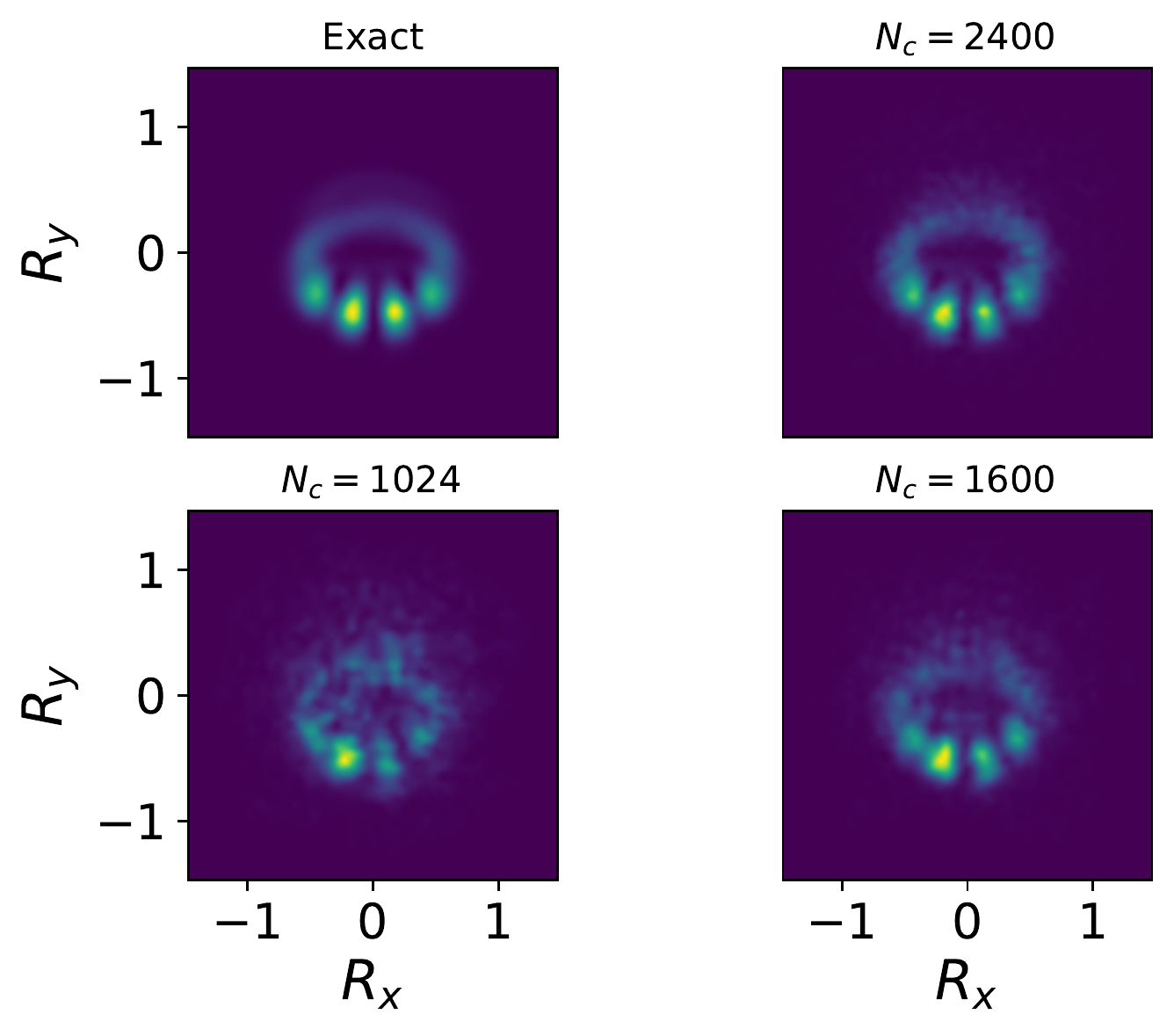}
 \caption{Convergece of the interference pattern arising from the Berry curvature with respect to number of bases elements, $N_c$. The computational time for each fourth-order Runge-Kutta timestep scales as $t=\mathcal{O}\left((N_c)^{a}\right)$ for $a=1.59\pm 0.06$.
 }\label{conical_intersection_convergence}
\end{figure}
We discuss here some of the technical details of the Berry phase intereference calculation demonstrated in ~\sref{Berry}. As in Ref.~\cite{schaupp2019classical} we took an electronic spatial grid from -12 to 12 a.u. with 81 grid points and a nuclear grid from -1.5 to 1.5 a.u. with 51 grid points alongside a time step of $dt=0.02$a.u. The initial wavefunction was constructed on this grid and the exact dynamics were propagated directly using a fourth-order Runge-Kutta integrator. 

The time-resolved picture presents this problem as a fully non-equilibrium problem, where the system starts already in a non-steady state, and so, the imaginary time sta-ICWF cannot be applied here to prepare the initial wavefunction. Instead, we stochastically sample the initial probability density $|\Psi_0(\bs r,\bs R)|^2$ with $\text{N}_c$ trajectories $\{\bs r^\alpha(0),\bs R^\alpha(0)\}$ that are used to construct CWFs $\phi_{r}^{\alpha}(\bs r,0)$ and $\phi_{R}^{\alpha}(\bs R,0)$ as defined in \eref{CWF1}. In this process, we respected the symmetry of the underlying initial state by symmeterizing the initial particle placement (and thereby complementarily symmetric slice CWFs) around the $R_y,\ r_y$ axes, meaning for each particle $R^{\alpha} = (R_{x}^{\alpha},R_y^{\alpha})$ we set $R^{\alpha+1}=(-R_{x}^{\alpha},R_y^{\alpha})$. 

In the dyn-ICWF, the pseudoinverse tolerance for $\mathbb{S}$ was set to $10^{-8}$ and the evaluation matrix elements of the electron-nuclear interaction potential term of \eref{4D_SM_W}, 
\begin{equation}
    \mathbb{W}_{\alpha\beta} = \iint d\bs R d\bs r \phi^{\alpha *}(\bs r)\chi^{\alpha *}(\bs R)W_{en}\phi^{\beta}(\bs r)\chi^{\beta}(\bs r) ,
\end{equation}
was accelerated by using an SVD decomposition to break up the 4 index potential $W_{en}(r_x,r_y,R_x,R_y)$ into a sum over electronic and nuclear two index vectors:
\begin{equation}
    W_{en}(r_x,r_y,R_x,R_y) = \sum_{l=1}^{N_{\sigma}}\sigma_l u_l(r_x,r_y)v_l(R_x,R_y).
\end{equation}
By tossing out $\sigma_l<10^{-4}$ we found that we were able to retain the accuracy of this potential to within a numerically tolerable limit with a speed up in computation time at a factor between 3.6 and 4.3 depending on hardware. A cubic interpolation to a grid twice as fine was used to smooth the images of the nuclear density 

In \fref{conical_intersection}, we show the first and second excited BOPESs associated to the extended Shin-Metiu model introduced in \sref{Berry}. 
dyn-ICWF results for $\text{N}_c = \{1024,1600,2400\}$ are shown in \fref{conical_intersection_convergence}. 
Due to the finesse of the interference pattern and its fragility with respect to the symmetry of the problem, the number of CWFs required to reproduce quantitatively the exact dynamics is relatively high compared to previous examples in Sec.~\ref{impact_SI} and \ref{SM_SI}. And yet, note that whilst the $\text{N}_c = 1024$ result do not reproduce the interference pattern accurately, they do qualitatively capture the nuclear dynamics by avoiding the forbidden region surrounding the conical intersection. This is in contrast to the mean-field result (figure 7 of Ref.~\cite{schaupp2019classical}), which fails to capture this qualitative feature of the nuclear dynamics.





\end{document}